\documentclass[11pt,twocolumn]{spieman}  
\usepackage{amsmath,amsfonts,amssymb}
\usepackage{graphicx}
\usepackage{setspace}
\usepackage{tocloft}
\usepackage{units}
\usepackage{upgreek}
\usepackage{bbm}
\usepackage[font=footnotesize, aboveskip=.2cm, belowskip=-0.25cm]{caption}
\usepackage[symbol]{footmisc}

\usepackage{textcomp}
\usepackage[absolute,overlay]{textpos}
\usepackage{helvet}

\title{Structured light for ultrafast laser micro- and nanoprocessing}

\author[a,*]{Daniel Flamm}
\author[b]{Daniel G. Grossmann}
\author[b]{Marc Sailer}
\author[a]{Myriam Kaiser}
\author[a]{Felix Zimmermann}
\author[a]{Keyou Chen}
\author[a]{Michael Jenne}
\author[a]{Jonas Kleiner}
\author[b]{\mbox{Julian Hellstern}}
\author[b]{\mbox{Christoph Tillkorn}}
\author[b]{Dirk H. Sutter}
\author[a]{Malte Kumkar}
\affil[a]{TRUMPF Laser- und Systemtechnik GmbH, Johann-Maus-Str.\,2, 71254 Ditzingen, Germany}
\affil[b]{TRUMPF Laser GmbH, Aichhalder Str.\,39, 78713 Schramberg, Germany }

\cftpagenumbersoff{figure}
\cftpagenumbersoff{table} 
\begin{document} 
\maketitle

\begin{abstract}
The industrial maturity of ultrashort pulsed lasers has triggered the development of a plethora of material processing strategies. Recently, the combination of these remarkable temporal pulse properties with advanced structured light concepts has led to breakthroughs in the development of novel laser application methods, which will now gradually reach industrial environments. We review the efficient generation of customized focus distributions from the near infrared down to the deep ultraviolet, e.g., based on non-diffracting beams and 3D-beam splitters, and demonstrate their impact for micro- and nanomachining of a wide range of materials. In the beam shaping concepts presented, special attention was paid to suitability for both high energies and high powers.
\end{abstract}

\begin{spacing}{1}


{\noindent \footnotesize\textbf{*}Address all correspondence to Daniel Flamm:\\ daniel.flamm@trumpf.com}


\section{Introduction}
\label{sect:intro} 
\begin{textblock*}{18cm}(1.7cm,1cm) 
   \centering
  \footnotesize \textsf{D.~Flamm \textit{et al.}, "Structured light for ultrafast laser micro- and nanoprocessing," Opt. Eng. 60(2) 025105 (24 February 2021); \url{https://doi.org/10.1117/1.OE.60.2.025105}.}
\end{textblock*}
\begin{textblock*}{19cm}(1.25cm,26.65cm) 
   \centering \footnotesize 
   \textsf{
   © 2021 Society of Photo‑Optical Instrumentation Engineers (SPIE). One print or electronic copy may be made for personal use only. Systematic reproduction and distribution, duplication of any material in this publication for a fee or for commercial purposes, and modification of the contents of the publication are prohibited. \url{https://doi.org/10.1117/1.OE.60.2.025105}.}
\end{textblock*}
Miniaturization---the trend to manufacture ever smaller products and components while retaining their function and quality---is a process that sooner or later affects all established technologies. The prospect of products with reduced sizes and weights and, therefore, costs, leads manufacturers to develop novel tools and processing strategies. Ultrashort pulsed lasers represent unique, industry-ready tools for machining in the micro- and nanometer scale. Pulse durations from $\sim\unit[100]{fs}$ up to $\sim \unit[10]{ps}$ and corresponding extreme peak intensities lead to interaction processes with all conceivable materials, independent of the state of matter or absorption behavior\cite{feng1997theory, nolte1997ablation, couairon2007femtosecond}. \par
Transparent and brittle materials, such as glasses or sapphire, can be found in almost every optoelectronic device and represent particularly challenging examples for processing mainly due to extreme mechanical or chemical properties. Especially here, ultrashort pulsed lasers show enormous potential, acting as subtle tools for a controlled energy deposition at the surface or inside the volume. At the same time, desired optical properties or implemented optical functionalities of non-processed, adjacent areas remain unaffected due to a marginal thermal diffusion\cite{chichkov1996femtosecond, nolte1997ablation}. A single, nearly diffraction limited Gaussian focus distribution usually provided by this class of lasers represents the ideal focus form for a certain laser machining process only in exceptional cases. More sophisticated spatial distributions become especially attractive when considering efficient, industry-compatible concepts intended to yield high throughputs\cite{Tillkorn2018,Flamm2019,jenne2020faci, pang2020focal}. Thus, in addition to the remarkable temporal properties of ultrashort pulsed lasers there is a need for tailored focus distributions in all spatial degrees of freedom \cite{Kumkar2014, Flamm2015, Flamm2019}. \par
\begin{figure*}
    \centering
    \includegraphics[width=1\textwidth]{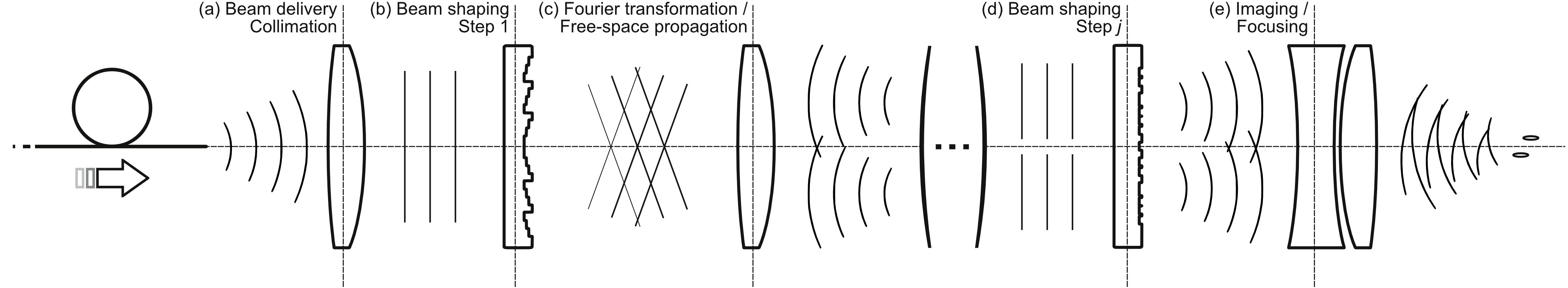}
    \caption{Modular concept for ultrashort pulse processing optics. Free space or fiber-guided pulses (a) are illuminating beam shaping elements (b) and propagate in free space or are transformed into their respective far-fields (c). After further possible beam shaping and propagation steps (d) the resulting optical fields are focused/imaged by an objective onto or into the workpiece (e). Alternating beam shaping and beam propagation/far-field transformation steps may be realized by single illuminating multiple elements or by illuminating single elements multiple times, cf. Sec.\,\ref{sec:drill}. Focusing/imaging units may be realized, for example, as $f$-$\theta$-objectives or as microscope objectives.\cite{Flamm2019}.}
    \label{sschool}
\end{figure*}

 
Recently, the term ``structured light'' has been used to cover advanced beam shaping concepts in which all spatial and temporal properties of laser radiation are manipulated to achieve tailored light states.\cite{rubinsztein2016roadmap, forbes2020structured}. In the present case, temporal characteristics are determined by the industrial ultrafast laser platforms used with band limited pulses of $\sim\unit[1]{ps}$ duration (or pulse trains thereof, \textcolor{black}{see, e.g. Refs.\,\citenum{herman2003burst,kerse2016ablation} for the beneficial use of burst modes}).\footnote{Simultaneous spatial and temporal focusing (SSTF) concepts\cite{Kammel2014} are not part of this review, since applied pulses exhibit durations of $>\unit[100]{fs}$.} The spatial properties, on the other hand, are determined by efficient and power-resistant concepts based on, e.g., diffractive-optical elements (DOEs)\cite{cumme2015regular, brodsky2019adjustable}, free-form optical elements (ROEs)\cite{wu2013freeform, cumme2015regular} or geometric phase holograms (GPHs)\cite{kim2015fabrication} (stationary or adaptive versions) illuminated single or multiple times. Together with the focusing unit (used NAs up to $0.5$)\cite{Flamm2019}, the beam shaping elements form the processing optics\cite{Flamm2019} usually fed with free-space or fiber-guided pulses \cite{wang2013hollow, baumbach2020hollow}, see Fig.\,\ref{sschool}.
\par 
The broad accessibility of liquid-crystal-based spatial light modulators has attracted the attention of the laser processing community, too \cite{mauclair2013ultrafast, Flamm2019, Jenne2018, jenne2020faci, Bergner2018, lazarev2019beyond}. There is ongoing progress regarding the display's ability to resist high peak and average powers, so that these devices may be part of future industrial processing optics forming ultrashort pulsed optical radiation in the kilowatt and tens-of-millijoules class\cite{sutter2019high, saraceno2019amazing, muller202010}.
\par Equally conceivable are concepts where the laser system itself is able to provide structured light. Such systems based, for example, on intra-cavity beam shaping\cite{naidoo2016controlled, forbes2017controlling}, amplified transverse fiber modes\cite{lin2020high, lin2020reconfigurable} or on coherent beam combining\cite{shekel202016kw, prossotowicz2020coherent,prossotowicz2020dynamic} enable complex focus shaping up to highly dynamic focus scanning and splitting. In the future, these concepts could replace the classical processing optics. Currently, however, the required degrees of freedom in focus distribution or the available (peak) power of the laser systems are still lacking.
\par 
The paper will review strategies for the micro- and nano-processing of transparent materials such as cutting (Sec.\,\ref{sec:cutt}), selective laser etching (Sec.\,\ref{sec:SLE}) or welding (Sec.\,\ref{sec:weld}) as well as surface ablation and drilling of metals (Sec.\,\ref{sec:drill}) that become possible or attractive for industry by spatiotemporal beam shaping. 

\section{Laser cutting of transparent materials}\label{sec:cutt}
There are different processing strategies known for cutting of transparent materials using ultrashort pulsed lasers\cite{nisar2013laser, Kumkar2014, kumkar2014cutting}. The concept presented in the following is based on a spatially controlled energy deposition into the volume of the workpiece.
\begin{figure*}
    \centering
    \includegraphics[width=1\textwidth]{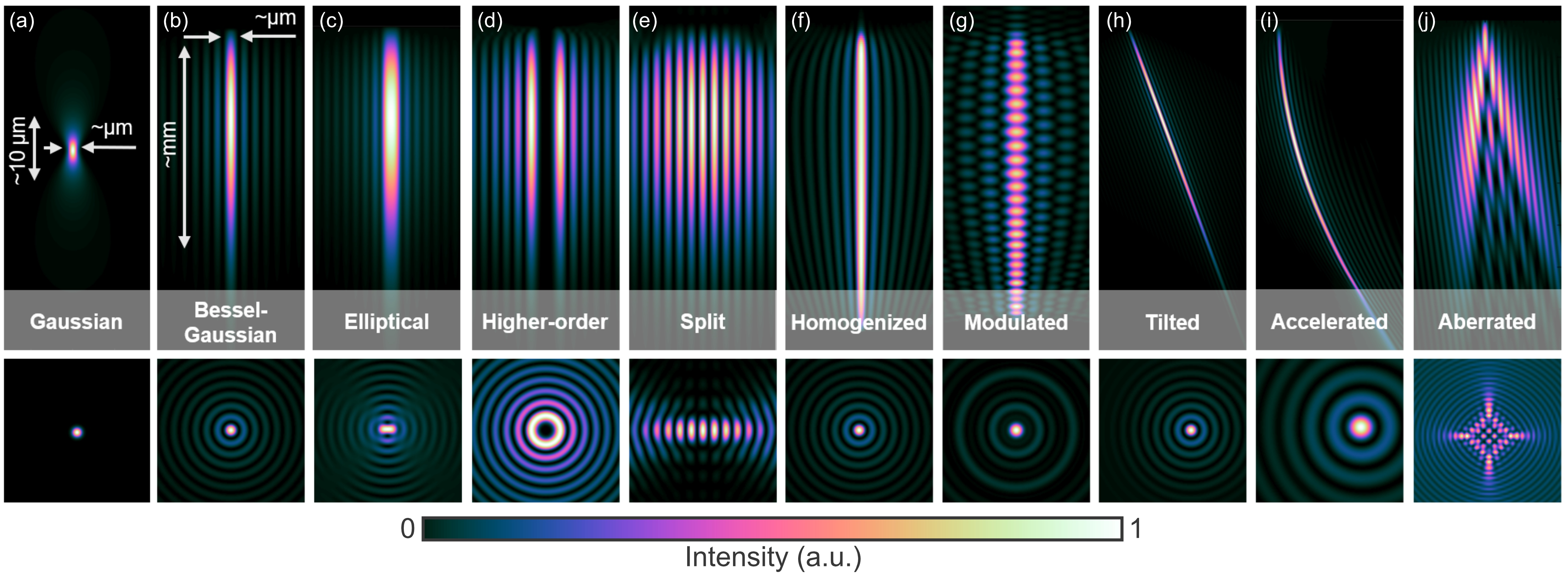}
    \caption{Menu of diffracting, non-diffracting and quasi non-diffracting beams designed for transparent materials processing\cite{Flamm2019}. All cases show the propagation behavior (cross section, top) and corresponding transverse intensity profile at a selected propagation distance (bottom). Gaussian focus distribution (for direct comparison) (a). Fundamental Bessel-Gaussian beam (b). Non-diffracting beams with elliptical central spots (c)\cite{chen2019generalized, jenne2020faci}, higher-order Bessel-Gaussian beams (d)\cite{vasilyeu2009generating, Flamm2017, xie2015tubular}, multi-focus distributions (e)\cite{chen2019generalized}, quasi non-diffracting beams exhibiting longitudinal homogenization (f)\cite{Flamm2015}, ``bottle-beams'' with on-axis modulation as superposition of quasi non-diffracting beams (g)\cite{Du2014, Flamm2015}, and propagation along tilted (h)\cite{Jenne2018a} or accelerating trajectories, also called second type of non-diffracting beams  (i)\cite{woerdemann2012structured, baumgartl2008optically,Chremmos2012, mathis2012micromachining}, as well as an example of an aberrated Bessel-like beam\cite{Jenne2018}. Here and in some of the following figures, we make use of David Green's color scheme representation.\cite{Green2011}}
    \label{bess00}
\end{figure*}
Induced micro explosions cause weakened areas characterized by voids, cavities and increased stress\cite{glezer1997ultrafast}.
\begin{figure}
    \centering
    \includegraphics[width=.48\textwidth]{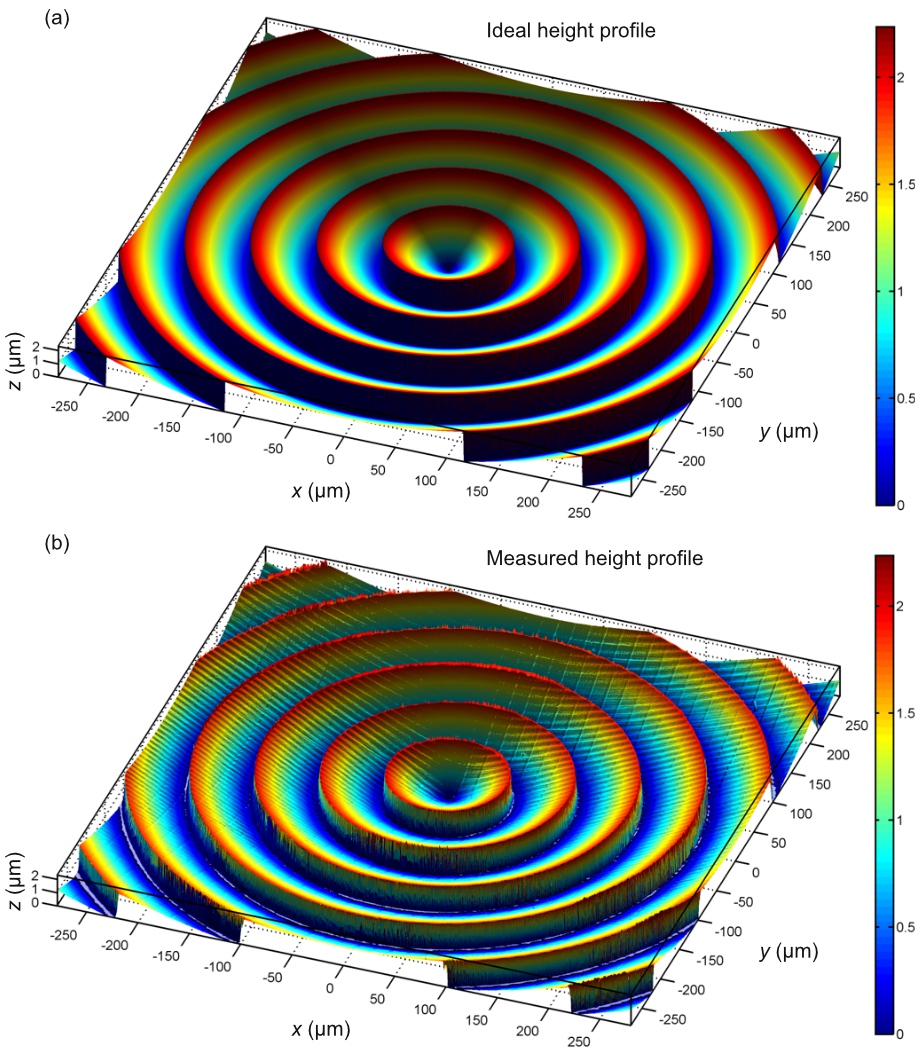}
    \caption{Examples of diffractive inverse axicons (central details). Ideal height profile (a) and  height profile measurement of a corresponding element fabricated via grey-tone lithography in fused silica using a laser scanning microscope (b). Here, the diffraction efficiency is above $\unit[99]{\%}$ with a zero-order power amount of $<\unit[0.5]{\%}$\cite{Flamm2019}. Please note the almost perfect axicon tip. \textcolor{black}{This diffractive solution represents a prime example for the realization of efficient and power resistant phase-only beam shaping elements shown throughout the manuscript as phase modulations, see e.g. Figs.\,\ref{bess} and \ref{fig:3dfoc}.}}
    \label{czj}
\end{figure}
A second process step applies external stress (e.g., thermal, mechanical, chemical) to these areas acting as breaking layer. The first step of the concept is achieved particularly efficient using an optical setup and a corresponding laser platform allowing to modify the entire substrate thickness with a single pulse. For this purpose, sufficient pulse energy is required on the one hand, and on the other hand a focusing optic, which creates an elongated focus distribution. Bessel-like beams represent a remarkable beam class for the controlled deposition of energy in a certain working volume\cite{McGloin2005, Flamm2015, bhuyan2010high}. They come to the fore with aspect ratios of longitudinal to transverse dimensions than can easily reach factors of higher than $10000$\cite{McGloin2005}, see Fig.\,\ref{bess00}\,(a), (b). Thus, the exact length of the processing beam can be adapted to the thickness of the transparent material enabling single-pass processing strategies\cite{Kumkar2014} for glasses of thicknesses $>\unit[10]{mm}$\cite{Flamm2019}. Further remarkable properties should only be mentioned here in brief: self-healing\cite{McGloin2005}, simple and efficient generation\cite{McGloin2005} and---particular interesting for the processing of transparent materials---a natural resistance to spherical aberrations\cite{Jenne2018, Flamm2019}. Figure \ref{bess00} provides a menu of non-diffracting beams tailored for various processing strategies \cite{Jenne2018a, Flamm2019, jenne2020faci}. \par 
First reports on Bessel-like beam generation were based on ring-slit apertures (far-field generation)\cite{Durnin1987}. However, for efficiency reasons, the axicon-based generation (near-field)\cite{McGloin2005} is preferred. An axicon is a conically ground lens and, by design, completely defined by axicon angle $\gamma$ and refractive index $n$. Using holographic axicons (both, digital and conventional) non-diffractive beams can be generated in very high qualities, as there arise no problems with manufacturing the axicon tip\cite{Brzobohaty2008, Flamm2019}, see diffractive element depicted in Fig.\,\ref{czj}. Here, a radial-symmetric transmission according to $T^{\text{axi}}\left(r\right)=\exp{\left[\imath\Phi^{\text{axi}}\left(r\right)\right]}=\exp{\left(\imath k_r r\right)}$ needs to be realized. In thin element approximation the radial component of the wavevector $k_r$ allows to directly assign a holographic axicon to its refractive counterpart via $k_r = 2\uppi \left(n-1\right)\gamma\lambda$ \cite{Leach2006, Flamm2019}. Illumination of these holograms by fundamental Gaussian beams generates Bessel-Gaussian beams of zero order. It is well known that minor modifications of $T^{\text{axi}}$ such as the multiplexing of phase vortices or $\uppi$-phase jumps allow the generation of Bessel-like beams of higher-order or superposition thereof \cite{vasilyeu2009generating, Bergner2018a, Flamm2019}. Here, the required phase modulation $\Phi$ is no longer radial symmetric but exhibits additional azimuthal dependencies such as, e.g., constant azimuthal slopes $\Phi\left(r, \phi\right) = k_rr+\ell\phi, \ell \in \mathbb{Z}$ \cite{vasilyeu2009generating, Flamm2019}. \par 
We use a generalization of this concept to generate non-diffracting beams with arbitrary transverse intensity profiles\cite{chen2019generalized, jenne2020faci}. Now, the generalized axicon-like phase transmission $T^{\text{gen}} = \exp{\left(\imath\Phi^{\text{gen}}\right)}$ is characterized by a phase modulation with arbitrary azimuthal dependencies $\Theta\left(\phi\right)$ and, thus,
\begin{gather}
\Phi^{\text{gen}}\left(r,\phi\right) = k_rr+\Theta\left(\phi\right).
\end{gather}
In principle, there are no restrictions regarding the exact dependency of $\Theta\left(\phi\right)$\cite{chen2019generalized, flamm2020generalized}. Continuous functions are as conceivable as those containing discrete jumps. The beam profiles created in this way are characterized by an enormous diversity regarding the transverse intensity profile while remaining almost all well-known beneficial features of Bessel-like beams ensured by the constant radial slope $\nicefrac{\partial\Phi^{\text{gen}}}{\partial r}=k_r$ given by the holographic axicon carrier.\cite{McGloin2005,jenne2020faci,chen2019generalized} 
\begin{figure}
    \centering
    \includegraphics[width=.48\textwidth]{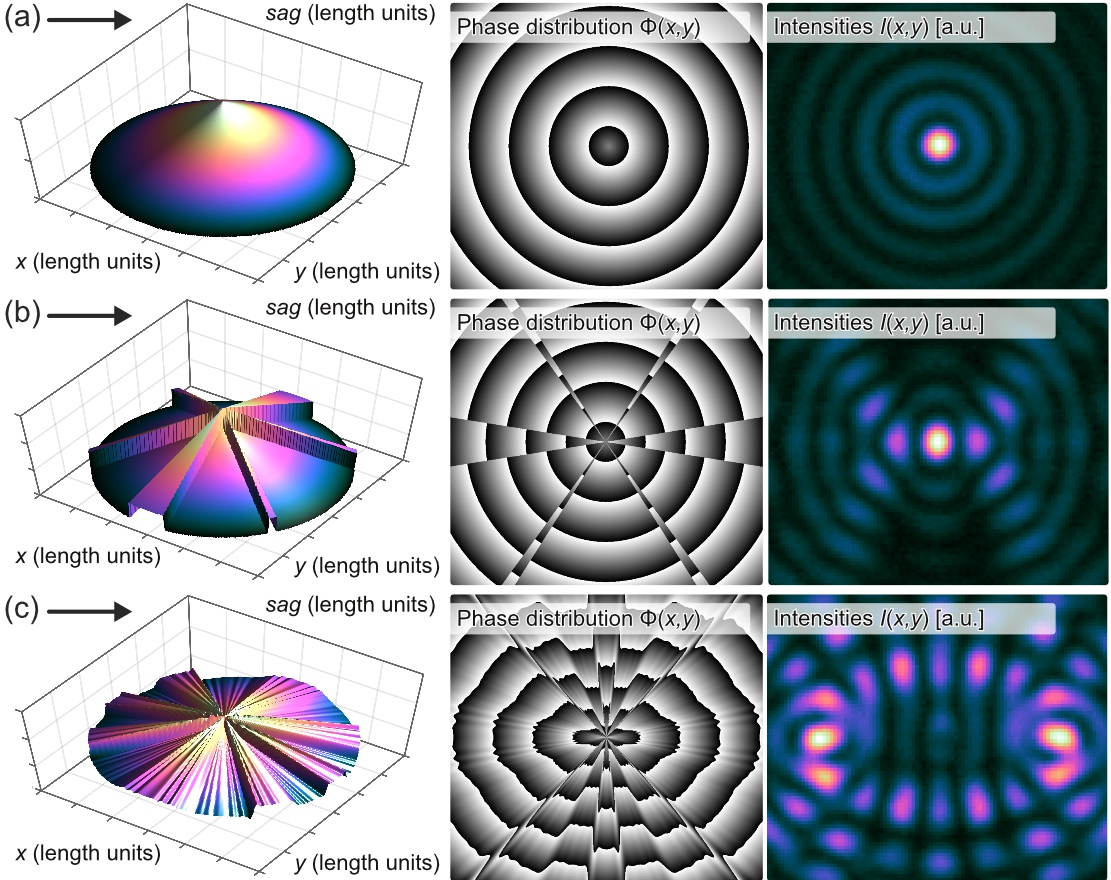}
    \caption{Shaping of non-diffracting beams with generalized axicons\cite{chen2019generalized}. For each example the sag profile of a refractive axicon (left), its holographic counterpart as phase-only transmission function (middle) and measured transverse intensity profile is depicted (right). Classical axicon shaping the fundamental Bessel-Gaussian beam (a). Generalized axicon generating an elliptical Bessel-like beam (b). Generalized axicon for shaping a non-diffracting beam exhibiting two spatially separated, elliptical intensity maxima (c). \cite{chen2019generalized, jenne2020faci}}
    \label{bess}
\end{figure}
Examples are provided in Fig.\,\ref{bess} where the measured transverse intensity profile of the fundamental Bessel-Gaussian beam (a) can directly be compared to an elliptical (b) and a twofold split non-diffracting beam (c), respectively, optically realized using generalized axicons. The implemented azimuthal fan segments, see $\Phi\left(x,y\right)$ in Fig.\,\ref{bess}\,(b), disturb the constructive on-axis interference of the fundamental Bessel-like beam resulting in an elliptical on-axis intensity maximum\cite{chen2019generalized}. For the particular case shown in Fig.\,\ref{bess}\,(b), the sag difference of the azimuthal segments results in azimuthal phase jumps of $\uppi$ which corresponds to a height difference of $\unit[1]{\upmu m}$ (at $\lambda = \unit[1]{\upmu m}$ wavelength) when realized as refractive version. For this reason, we prefer the diffractive realization, similar to the example depicted in Fig.\,\ref{czj}\,(b).\cite{Flamm2019} 
\par 
Throughout this work several phase profiles are shown, cf. Figs. \ref{bess},\,\ref{fig:3dfoc}, for the realization of which there are various options. One variant is based on an etched microstructure in fused silica, see example depicted in Fig.\,\ref{czj}. Particular efficient DOEs may be achieved by applying an adapted anti-reflection nano-texture\cite{hobbs2013contamination} exhibiting pulsed laser damage resistances which are typically several factors higher than conventional thin-film anti-reflection coatings and, thus, close to the thresholds of the unprocessed surface\cite{hobbs2007high}. Diffractive elements similar to the one depicted in Fig.\,\ref{czj} have been subjected to long-term ultrashort laser pulse high-power and high-energy tests. For pulse durations of $\sim\unit[1]{ps}$ we applied average powers above $\unit[1]{kW}$ as well as pulses in the order of tens-of-millijoules\cite{dietz2020ultrafast}. At resulting fluences of $\sim\unit[0.1]{J/cm^2}$ and peak intensities of $\sim\unit[1 \times 10^{11}]{W/cm^2}$ no signs of degradation at the DOE's surface or optical functionality were apparent.
\par
The realization of the concept as processing optics is explained by means of Fig.\,\ref{sschool}. The collimated raw beam is illuminating the axicon-like DOE [Fig.\,\ref{sschool}\,(b)]. A subsequent telescopic setup is imaging the non-diffracting beam into the workpiece with a well-defined (de-)magnification factor [Fig.\,\ref{sschool}\,(c),\,(e)]. For this purpose, long-working distance microscope objectives with NAs up to $0.5$ are available.\cite{Flamm2019, grossmann2020scaling}.
\par
\begin{figure}
    \centering
    \includegraphics[width=.48\textwidth]{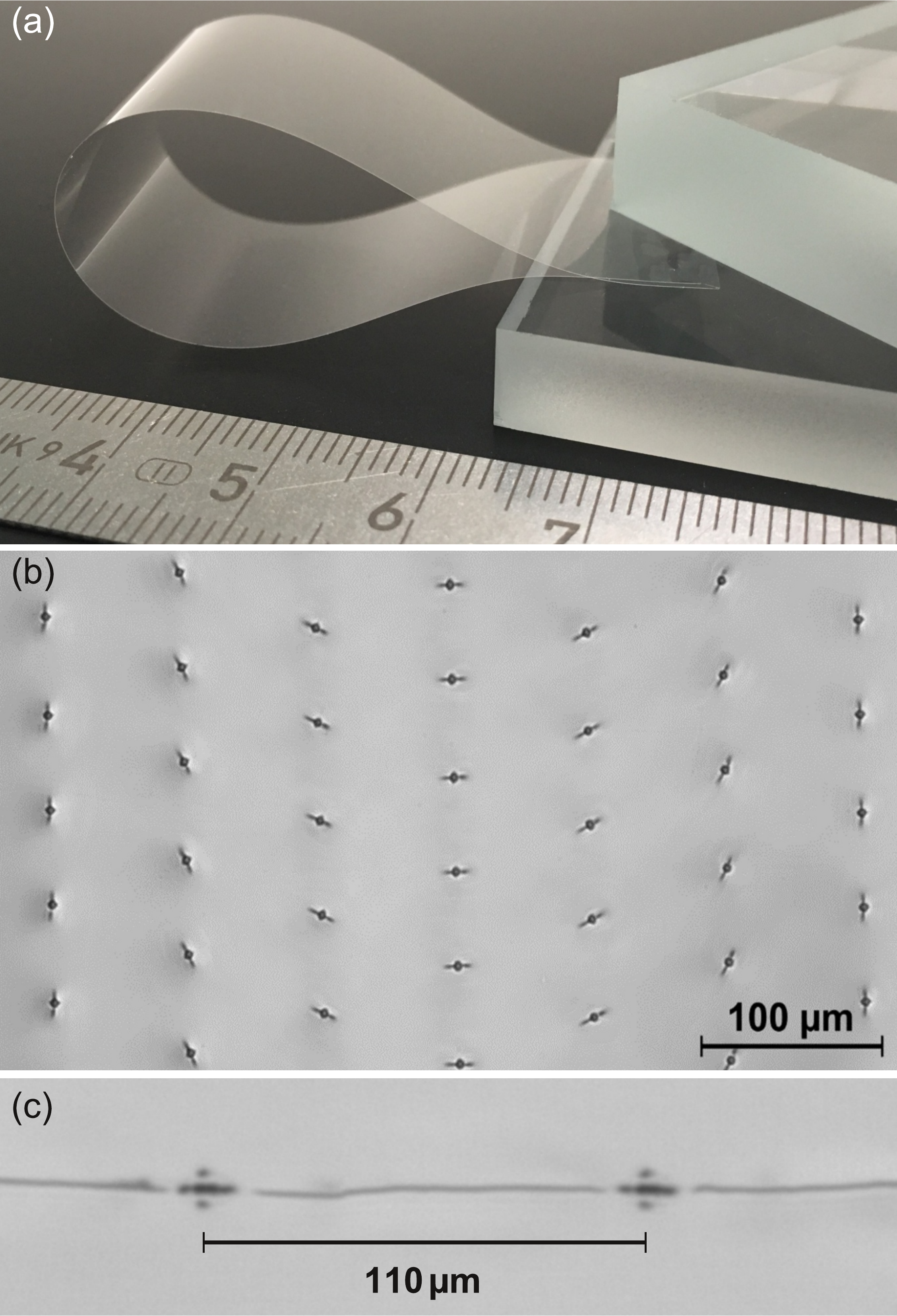}
    \caption{Processing results of facilitated separation of ultra thin ($<\unit[0.05]{mm}$) and thick glass substrates ($>\unit[10]{mm}$) (a). For the latter case the TOP Cleave cutting optics and a TruMicro Series 5000 laser source providing pulse energies of $\unit[1.5]{mJ}$ was used. Permanent material modifications at the glass surface from single ultra short pulses spatially shaped as elliptical non-diffracting beam, similar to the example depicted in Fig.\,\ref{bess}\,(c). The orientation of cracks is controlled by rotating the central beam shaping element in $30^{\circ}$ steps (b)\cite{chen2019generalized, jenne2020faci}. Modifications directly connected by cracks may well be $\unit[100]{\upmu m}$ apart (c). \textcolor{black}{So far, the concept for full-thickness crack orientation control by rotating asymmetric elongated beam profiles is restricted to amorphous materials.}}
    \label{fig:faci}
\end{figure}

As mentioned in the beginning of this section, the modification step is only one part of the entire cutting process. The actual separation process remains challenging especially if comparatively thin $<\unit[0.3]{mm}$ and thick glasses $>\unit[2]{mm}$ need to be separated. In the first case, already smallest deviations from the laser induced modification layer, such as, e.g., local micro cracks, are sufficient to separate the material not along the intended contour. In the second case, for mm-scaled glass substrates, the required force necessary for separation exceeds $\unit[1]{kN}$\cite{jenne2020faci} which makes controlled cutting with high quality equally difficult. For these reasons, very recently, there is ongoing research for facilitated cutting of glasses using asymmetric Bessel-like beams to control the formation of cracks. Tailored transverse intensity profiles of non-diffracting beams can cause modifications with preferential direction reducing the necessary breaking force. Established concepts make use of spatial frequency filtering or controlled aberrations \cite{hendricks2016, Meyer2017, Dudutis2018} but come at the expense of significant power loss or reduced process stability. Our concept using generalized axicons which allows to shape non-diffracting beams with an elliptical on-axis intensity profile, cf.~Fig.\,\ref{bess}, represents a promising approach to further reduce the necessary breaking force of thick glass substrates and to control the separation along predefined contours of ultra-thin glasses, respectively. More details about this facilitated glass separation concept are provided in Ref.\,\citenum{jenne2020faci}. Selected processing results are depicted in Fig.\,\ref{fig:faci} highlighting the successful glass cutting for two extreme glass substrate geometries.
\par
\begin{figure}
    \centering
    \includegraphics[width=.48\textwidth]{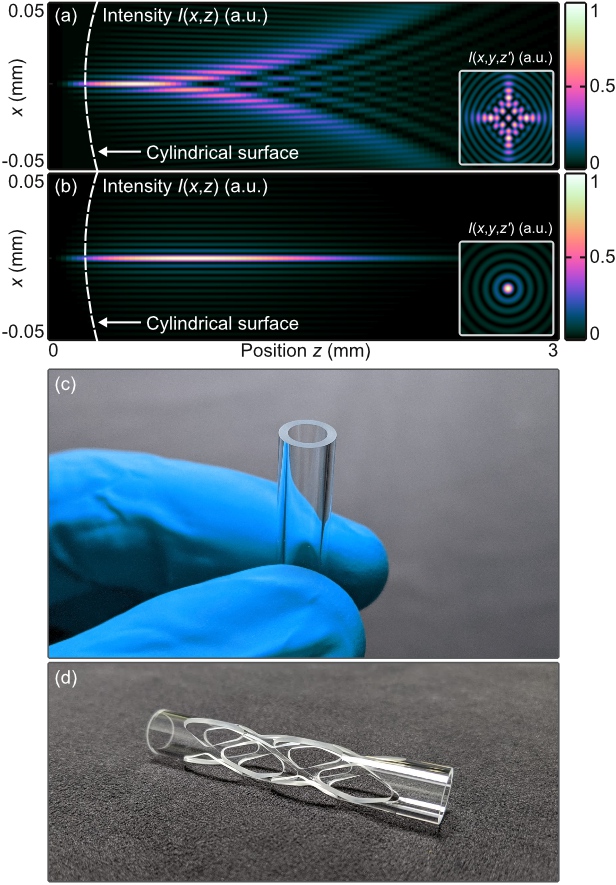}
    \caption{Cutting of glass tubes with complex inner and outer contours. The aberrated (a) and aberration-corrected (b) elongated beam profile is depicted in terms of simulated intensity cross-sections $I\left(x,z\right)$ and $I\left(x,y\right)$ (see subsets), respectively. The cylindrical surfaces (radius, z-position) are sketched only schematically. A processed glass tube (diameter $\sim\unit[10]{mm}$, wall thickness $\sim\unit[1]{mm}$) with mechanical outer contour separation is depicted in (c). The complex inner contours of the sample depicted in (d) were achieved using a selective etching strategy, cf. Sec.\,\ref{sec:SLE}. Processing results were achieved using a modified TOP Cleave cutting optics in combination with $\unit[100]{\upmu J}$-pulses emerged from a TruMicro 2000 series laser.}
    \label{fig:zaxicon}
\end{figure}

Based on the developed processing strategies for perpendicular glass cutting, recently, there is an increased demand for customized glass edges, such as, e.g., chamfer or bevel structures. Apart from visual benefits, this is due to the increased edge stability and a reduction of potential edge fractures. Additionally, substrate geometries are becoming more complex and there is interest in cutting glass with curved surfaces, such as tubes, syringes and ampoules as required in large quantities in medical technology. The challenge for both cases is that the applied non-diffracting focus distributions require sensitive aberration corrections to compensate phase disturbances caused by the beam transition through tilted or curved interfaces. Figure \ref{fig:zaxicon} illustrates this concept (for details, see Refs.\,\citenum{Jenne2018, Jenne2018a}) at the example of processing glass tubes with given diameter.
The propagation of a Bessel-Gaussian beam behind a cylindrical glass surface is simulated in Fig.\,\ref{fig:zaxicon}\,(a) where a complex interference pattern is obtained with reduced peak intensities especially for larger propagation distances $z$. Now, we virtually apply the aberration correction and achieve the almost undisturbed profile of the well-known Bessel-Gaussian beam, see Fig.\,\ref{fig:zaxicon}\,(b). Processing samples are depicted in (c),\,(d) proving the efficacy of the concept for single-pass cutting of glass tubes of diameter $>\unit[1]{mm}$ with complex inner and outer contours. \textcolor{black}{Structured light concepts in this and the following sections (Secs.\,\ref{sec:SLE} and \ref{sec:weld}) were applied to transparent materials (glasses and crystals) with refractive indices below $2$ at the laser processing wavelength. For processing high refractive index transparent materials ($n>2$) such as, e.g., silicon, silicon carbide or diamond, the processing strategies presented, for example with regard to single-pass cutting using non-diffracting beams, cannot be applied as is. Significantly reduced processing thresholds at the surface as well as inside the volume prevent a spatially controlled energy deposition. Comparatively low intensities already lead to surface ablation perturbing the propagation of optical fields into the volume. The handling of the high index mismatch is equally challenging for the optical design as the strong refraction and Fresnel reflections prevent certain beam shaping concepts. However, space-time structured light concepts \cite{Kammel2014}, especially techniques providing radially chirped simultaneous spatial and temporal focusing\cite{meier2015application} represent promising potential solutions. Radial symmetric grating structures, similar to the one depicted in Fig.\,\ref{czj} but with higher spatial frequencies, generate a well-defined radial dispersion resulting in Gaussian or Bessel-Gaussian field distributions exhibiting short pulses and, thus, high intensities only in the focal region where energy is intended to be deposited. In the areas located immediately before or after the spatial focus, pulse durations are significantly longer and intensities as well as unwanted material defects are reduced.}

\section{Selective laser etching of transparent materials}\label{sec:SLE}
The combination of ultrashort pulsed volume modifications and subsequent selective chemical etching of dielectrics, known as selective laser etching (SLE), enables rapid fabrication of 3D glass structures of arbitrary shape with smallest structural features down to the $\unit[10]{\upmu m}$ scale\cite{wortmann2008micro, hermans2014selective, Kumkar2016}. The remarkable 3D fused silica articles provided by, e.g., LightFab\cite{LF3dptint} convince with extreme geometries, including micro tunnels, cavities or mounted moving parts. Usually, the corresponding processing strategy is based on 3D-scanning a single Gaussian focus distribution into the transparent workpiece to spatially control the nonlinear energy deposition. The concept presented in the following focuses on the efficient fabrication of high-quality 2D geometries ranging from mm-scaled complex inner contours down to nano-channels in transparent materials. The key to the efficient and industry-ready processing concept is to generate selectively etchable elongated modifications connected by cracks provided by spatially shaped ultrashort pulses using the concepts on non-diffracting beams presented in Sec.\,\ref{sec:cutt}. The employed beam profiles thus generated are length adapted by the processing optics allowing for full-thickness, single-pass modifications of the substrate. Our approach is thus remarkable efficient, allows for throughput scaling and to exploit the complete pulse energy performance of the respective laser platform \cite{Flamm2019}.\par 
\begin{figure}
    \centering
    \includegraphics[width=0.48\textwidth]{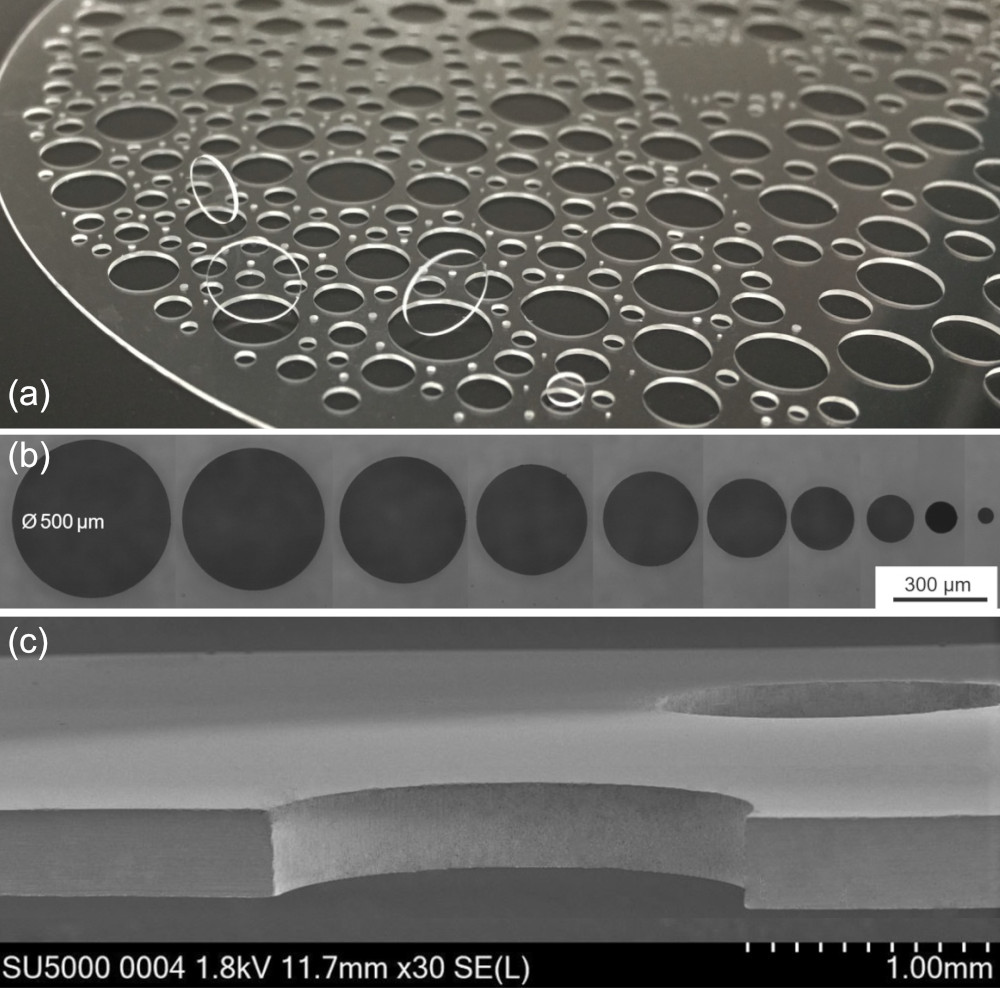}
    \caption{Fused silica wafer ($\unit[4]{in}$) with thickness of $\unit[350]{\upmu m}$ with high-quality glass vias of diameters between $\unit[0.25]{mm}$ and $\unit[6]{mm}$ fabricated by selective laser etching (a). Microscope images of the laser entrance glass side confirm high-quality edge contours (b). Taper-free glass vias proven by SEM image (c)\cite{flamm2020structured}.}
    \label{fig:yakuv}
\end{figure}
During the laser process, elongated modifications, which extend from the beam entrance to the exit side of the material to be processed, are arranged sequentially along a contour. The modified volume shows typically a $100$ times higher etching rate compared to the base material. Depending on the respective material to be processed, adapted pulse energies, durations and a controlled spatio-temporal energy deposition strategy enable the fabrication of high-quality, taper-free 2D-geometries with high etching rates. Figure \ref{fig:yakuv}\,(a) depicts a processed fused silica wafer of $\unit[350]{\upmu m}$ thickness into which high-density through glass vias (TGVs) of different diameters (between $\unit[0.25]{mm}$ and $\unit[6]{mm}$) were etched. By applying $30$\,wt.-\% KOH etch solutions in an ultrasonic bath at a temperature of $\unit[80]{^\circ C}$ etching rates of about $\unit[20]{\upmu m/min}$ were achieved.
\par 
Besides an increase of throughput there is a demand for further miniaturization of the etched geometries. Using a non-diffracting beam in combination with the single pass contour modification strategy described above, allows to generate crack-free microholes down to $\unit[50]{\upmu m}$ in diameter, see Fig.\,\ref{fig:yakuv}\,(b). The smallest possible aspect ratio of hole diameter to material thickness is about $1:10$---almost independent of the processed transparent material. We provide an SEM image of a cut TGV profile proving that there is no taper angle at hand, see Fig.\,\ref{fig:yakuv}\,(c).
\begin{figure}
    \centering
    \includegraphics[width=0.48\textwidth]{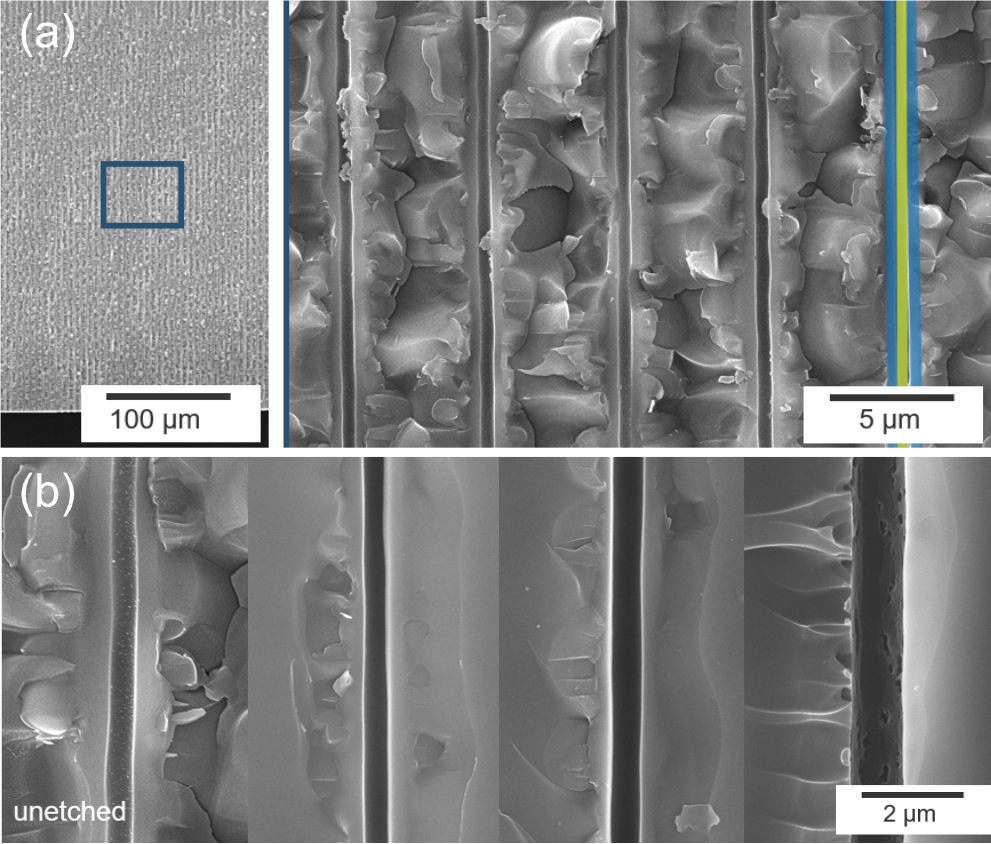}
    \caption{Nanochannels in fused silica fabricated by single-pulse modifications from a non-diffracting beam and ultrashort pulses. The high aspect channels exhibit diameters of $\unit[500]{nm}$ without visible taper angles and without applying subsequent etching (a). An additional etch step enables to increase the channel diameter. From left channel to right: no etching, $\unit[5]{min}$, $\unit[10]{min}$, and $\unit[15]{min}$ etching in $30$\,wt.-\% KOH etch solution (b).\cite{flamm2020structured}}
    \label{fig:nannchnn}
\end{figure}
\par
In addition to the presented complex contours and microholes another innovation in laser processing is the generation of hollow nanoscale channels in glass and sapphire. The processing strategy is again based on using a non-diffracting beam and a TruMicro laser platform enabling the formation of hollow modifications in glasses using a single pulse only, without subsequent etch steps. Depending on optical configuration and pulse specifications, nanochannels of desired length and diameter can be achieved. 
In Fig.\,\ref{fig:nannchnn}\,(a) nanochannels along a glass thickness of $\unit[350]{\upmu m}$ are shown. The high aspect channels exhibit diameters of $\unit[500]{nm}$ without visible taper angles. In the case of glass, a fine adjustment of the nanochannel’s diameter is possible by an additional selective etching step. For example, as shown in Fig.\,\ref{fig:nannchnn}\,(b) a short etching time in KOH solution allows the channels to be expanded to a diameter of $\unit[1]{\upmu m}$.
\begin{figure}
    \centering
    \includegraphics[width=0.48\textwidth]{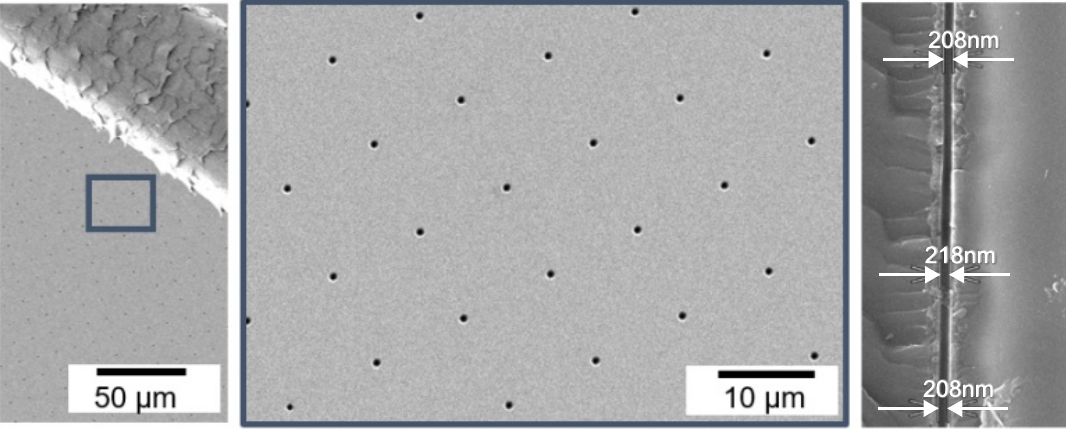}
    \caption{Scanning electron microscope image of an array of nanochannels etched into ultrafast laser modified sapphire. For size comparison, on the upper left, a human hair is microscoped.\cite{flamm2020structured}}
    \label{fig:nannchnn2}
\end{figure}
In the case of processing sapphire, elongated modifications filled with an amorphous phase instead of hollow nanochannels appear after the modification process.\cite{kaiser2019selective}. An additional selective etching step to open the amorphous phase is required resulting in $\unit[200]{nm}$-diameter hollow channels, see Fig.\,\ref{fig:nannchnn2}. \textcolor{black}{Therefore, controlling the channel diameter by varying etching parameters, similar to glass (cf.~Fig.\,\ref{fig:nannchnn}), is not straightforward for crystals.} We demonstrate highest reproducibility with a fabricated array of nanochannels that can be compared to the dimensions of a human hair (see left hand side in Fig.\,\ref{fig:nannchnn2}).
\begin{figure*}
    \centering
    \includegraphics[width=1\textwidth]{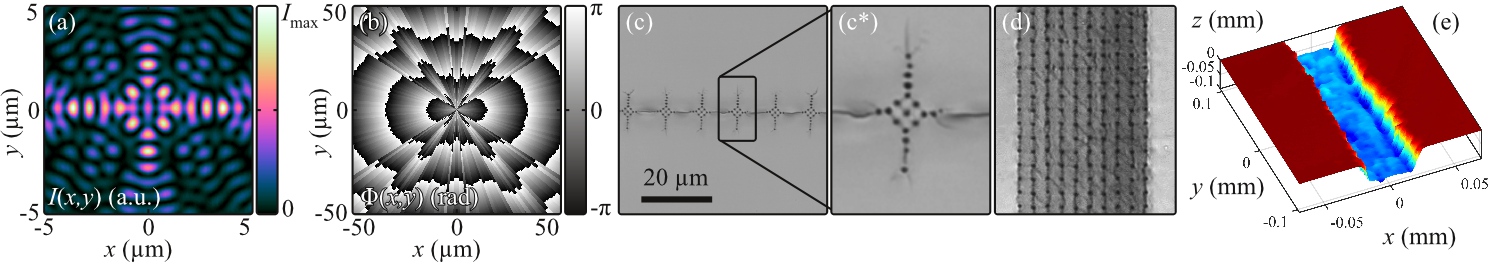}
\caption{Efficient SLE of large glass volumes based on spatial crack control. Cross-like transverse intensity profile of the non-diffracting beam (a) and phase modulation of the corresponding generalized axicon (b)\cite{chen2019generalized}. Microscope image of surface modifications with cross-like crack orientation (c), (c*). Plurality of surface modifications on a Cartesian grid connected by perpendicular running cracks (d). Resulting large-volume trench efficiently removed via selective etching of the crack-modified volume (e)\cite{kaiser2019selective}.}
    \label{fig:corrsss}
\end{figure*}
\par 
For the case that comparatively large volumes need to be etched efficiently, beam shaping again proves to be particularly useful. In Fig.\,\ref{fig:corrsss} a non-diffracting beam with cross-like transverse intensity distribution is depicted (a) that generates volume modifications with cross-like crack orientations (b). A plurality of such modifications on a Cartesian grid connected by perpendicular running cracks (d) allows to efficiently etch a large volume trench (e).
\par
At this section's end we would like to point out that the complex inner contours from glass tubes, see Fig.\,\ref{fig:zaxicon}\,(d), were removed using a selective etching strategy, too. Thus, discussed concept can be applied to curved glass articles.

\section{Laser welding of transparent materials}\label{sec:weld}
The short pulse duration and subsequent nonlinear absorption within the focal region in combination with heat accumulation from pulse to pulse enables local melting of transparent materials within a confined region that allows for permanent joining. Even though first reports of this technology were already published in 2001 by Tamaki \textit{et al.} \cite{tamaki2005welding} the technology is only gradually transferred to industrial applications so far. This is due to the high surface preparation requirements, i.e. low roughness and flatness in order to reduce potential gaps close to zero. On the other hand, the size of the gap that can be closed strongly depends on the glass properties that limit the weld seam dimension and laser power that can be applied. In glasses with low thermal diffusivity such as fused silica, thus large weld seams can be induced that allow for the bridging of gaps up to a few $\upmu$m \cite{richter2015toward, cvecek2015gap}. Moreover Chen \textit{et al.} reported on the formation of a melt pool using high-scan speeds which allows to accumulate melt and bridge significant larger gaps, see Ref.\,\citenum{chen2019picosecond}. However, crack formation inside the melt and reduced optical quality of the seams are obtained. \textcolor{black}{A detailed study about spatial and optical properties of weld seams including SEM images is provided in Ref.\,\citenum{richter2011bonding}.} \par
In our approach temporal and spatial beam shaping is used to overcome basic limitations while keeping the stability of the joints high (up to the volume material) and generate high-transparency seams. Details of the method can be found in Ref.\,\citenum{zimmermann2019situ}. The basic idea is to enlarge the laser-induced seam based on an elongated beam shape. By this the absorption region is distributed over an elongated region resulting in homogeneous melting during ongoing illumination. In addition, the induced stress is strongly reduced allowing for larger modifications and crack-free welding. \textcolor{black}{We refer to Jenne \textit{et al.}\cite{Jenne2018b} for a detailed study on the material's absorption and stress response to pulses shaped as Gaussian foci and elongated focus distributions, respectively.}
\begin{figure}
    \centering
    \includegraphics[width=0.43\textwidth]{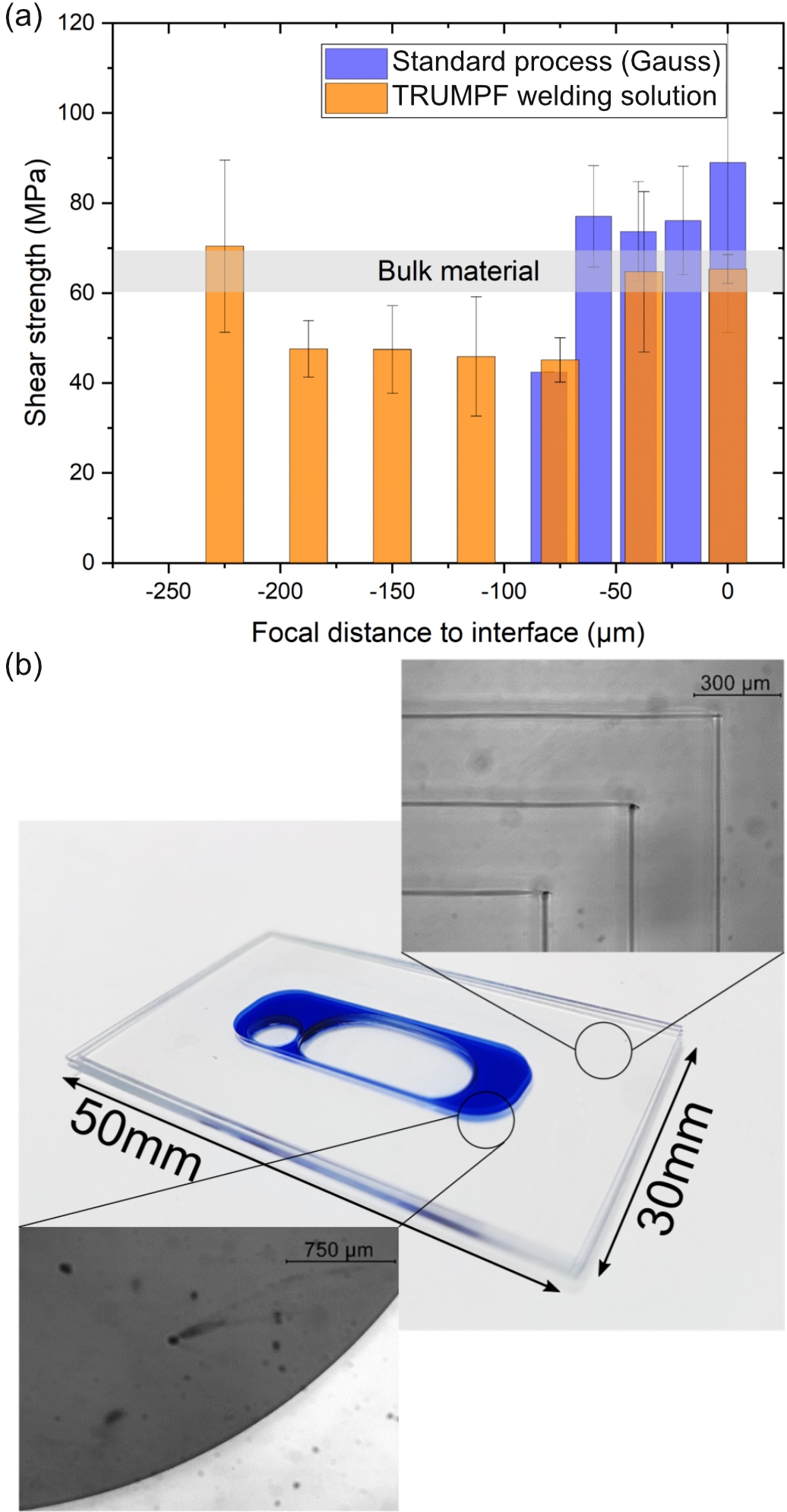}
    \caption{Measured shear strength of welded Gorilla glass (Corning) in dependence of the focal position relative to the interface for standard focusing and elongated beam shape (TOP Weld optics) (b). Encapsulated liquid (c) made of three individual glass layers whereas the middle glass exhibits a rectangle shaped cavity (fabricated by laser modifying and etching). The microscope images show the weld seams (above) and the etched contour (below).\cite{flamm2020structured}}
    \label{fig1}
\end{figure}
\par 
The TRUMPF ultrashort pulse welding portfolio consists of an ultrashort pulse laser source (TruMicro 2000 series) and the TOP Weld optics. The focus distribution thus generated leads to much higher process performance as shown by the measured shear strength in Fig.\,\ref{fig1}\,(a). The focal tolerance of about $\unit[230]{\upmu m}$ in $z$-direction is about three times higher compared to state-of-the-art focusing (Gaussian beam shape) while keeping the breaking stability as high as the volume material. 
\par 
By using temporal pulse energy modulation the weld performance can even be improved. As shown by Nakamura \textit{et al}. the temporal modulation of the pulse energy can reduce permanent stress in and around the weld seam and allows for larger seam dimensions \cite{nakamura2017suppression}. Adding this technique, which can directly be set in the laser control of the TruMicro 2000 series, the focal tolerance shown in Fig.\,\ref{fig1}\,(a) can be increased up to $\unit[300]{\upmu m}$ for Gorilla glass. Another important feature for welding in industrial environment is the bridging of gaps. As mentioned above state-of-the-art techniques are limited to about $\unit[3]{\upmu m}$. By using the TOP Weld optics gaps up to $\unit[7]{\upmu m}$, together with pulse energy modulation gaps even up to $\unit[10]{\upmu m}$ can be bridged. By this, real world applications in the field of consumer electronics, bio medicine or encapsulation for MEMS devices can be addressed.
\par 
One of the main challenges for encapsulation denotes the gas and liquid tightness of welded devices. To test the resulting tightness after welding protection caps of laser light cables were characterized (see Ref.\,\citenum{kaiser2016} for details). By this a hermeticity to the power of $-9$ could be measured confirming the feasibility for encapsulated devices such as microfluidics. This has further been demonstrated by encapsulating a blue liquid shown in Fig.\,\ref{fig1}\,(b). The article is made of three individual glass layers (each $\unit[30]{mm} \times \unit[50]{mm} \times \unit[0.55]{mm}$) whereas the middle glass exhibits a rectangle shaped cavity that has been fabricated by laser scribing (using the same laser setup) and etching (contour shown in the inset). Several weld seams (10) were inscribed in a parallel manner around the cavity (see inset). The encapsulated device showed high stability without leakage of the liquid. 
\par 
Since femtosecond laser pulse welding around $\unit[1]{\upmu m}$ wavelength of transparent materials requires moderate pulse energies $<\unit[10]{\upmu J}$\cite{tamaki2005welding} concepts for throughput scaling are becoming increasingly important\cite{Kumkar2016, Flamm2019,flamm2020structured}. This becomes even clearer if one considers the recent progress in pulse energy and average power of today's ultrafast laser platforms\cite{sutter2019high}. We use a structured light concept based on a diffractive 3D-beam splitter to discuss the potential of parallel processing for particularly efficient welding of transparent materials\cite{Kumkar2017, Flamm2019}. \par
One way to displace a focus from its original position behind a lens is achieved by introducing phase modifications into the wavefront of the illuminating optical field. In terms of Zernike polynomials\cite{Noll1976}, a proper choice of tip/tilt modes in combination with a defocus allows to control the transverse $\left(\Delta x, \Delta y\right)$ and longitudinal $\Delta z$ displacement, respectively. We extend this simple concept for the manipulation of a single focus to a 3D-beam splitting concept by exploiting the linearity property of optics and multiplex the corresponding holographic transmission functions\cite{Flamm2012, Forbes2016, silvennoinen2014parallel} (one for each focus to be placed in the working volume).

\begin{figure*}
    \centering
    \includegraphics[width=0.9\textwidth]{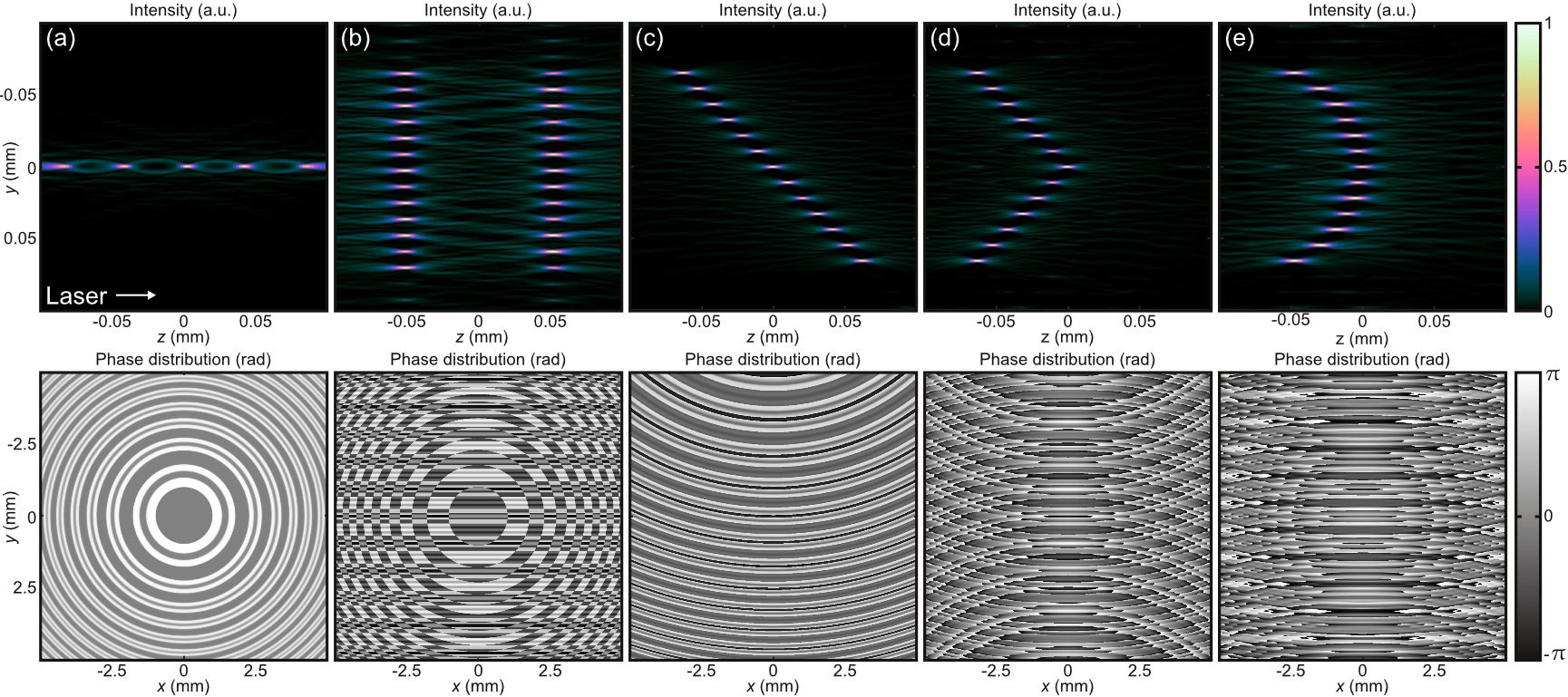}
    \caption{Selection of 3D-focus distributions [intensity cross sections, $I\left(x=0, y, z\right)$, top row, $z$-axis corresponds to propagation direction] with corresponding phase-only transmission functions ($\arg{\left[T^{\text{tot}}\left(x,y\right)\right]}$, bottom row) for beam generation. The Gaussian-foci are arbitrarily distributed within the working volume. Regular longitudinal focus arrangement (a). Combination of regular longitudinal and transversal beam splitting (b). Focus arrangement along a $45^{\circ}$-straight line (c). Focus arrangement along a $90^{\circ}$-angle (d). Focus arrangement along an accelerating trajectory (half circle) (e).\cite{Flamm2019}}
    \label{fig:3dfoc}
\end{figure*}

Transverse displacement for the focus of order $j$ is achieved by setting a linear blaze grating with spatial frequency $\mathbf{K}_j = \left( K_{x,j}, K_{y,j}\right)$ in the front focal plane of a lens with focal length $f_{\text{FL}}$. The corresponding transmission function then reads as 
\begin{equation}\label{eq:grat1}
	T_{j}^{\text{trans}}\left(\mathbf{r}\right) = \exp{\left[-\imath\left(\mathbf{K}_j\mathbf{r}+\phi_j\right)\right]}
\end{equation}
and yields a transverse displacement according to $\Delta x_j = f_{\text{FL}}\tan{\left[\sin^{-1}{\left(\lambda K_{x,j}/2\uppi\right)}\right]}$ (straightforward for $\Delta y_j$). Longitudinal displacement $\Delta z_j$ of the $j$th-order focus is realized by introducing the defocus ``aberration'' using, e.g., a holographic lens transmission with focal length $f_j$
\begin{equation}
T_{j}^{\text{long}}\left(\mathbf{r}\right) = \exp{\left[-\imath\uppi r^2/\left(\lambda f_j\right)\right]}.
\end{equation}
Then, the longitudinal shift is directly deduced from $\Delta z_j \sim -f_{\text{FL}}^2 / \left(f_{\text{FL}}+f_j\right)$. Combinations of both shifts are achieved by multiplying both transmissions $T_j = T_{j}^{\text{trans}}T_{j}^{\text{long}}$. Multiplexing these $j_{\text{max}}$ transmission functions will yield the total transmission according to
\begin{equation}\label{eq:tot}
T^{\text{tot}}\left(\mathbf{r}\right) = \sum_j^{j_{\text{max}}} T_{j} = \sum_j^{j_{\text{max}}} T_{j}^{\text{trans}}\left(\mathbf{r}\right)T_{j}^{\text{long}}\left(\mathbf{r}\right).
\end{equation}
In general, this multiplexing scheme will result in a complex valued transmission $T^{\text{tot}}\left(\mathbf{r}\right) = A^{\text{tot}}\left(\mathbf{r}\right) \times \exp{\left[\imath\Phi^{\text{tot}}\left(\mathbf{r}\right)\right]}$. Different approaches exist to realize such a transmission as phase-only hologram, see, e.g., Davis \textit{et al.~}\cite{Davis1999} or Arriz{\'o}n \textit{et al.~}\cite{Arrizon2007}. However, a particularly efficient and simple solution represents $A^{\text{tot}}\left(\mathbf{r}\right) = \mathbbm{1}$, thus setting the amplitude modulation to unity and directly use $T^{\text{tot}}\left(\mathbf{r}\right) =\exp{\left[\imath\Phi^{\text{tot}}\left(\mathbf{r}\right)\right]}$ as phase-only transmission. This approach will yield optical powers in unwanted diffraction orders, but, nonetheless, will be significantly more efficient than aforementioned phase-coding techniques. However, it has negative impact on the uniformity of the split spots. To restore equal power distribution a set of constant phase offsets $\left\{\phi_j\right\}$ in the grating representation of Eq.\,(\ref{eq:grat1}) can be found by an iterative optimization routine. Here, the optical field in the working volume and the optical power of the $j_{\text{max}}$ spots have to be simulated for each iteration\cite{Leutenegger2006}. This iterative Fourier-transform algorithm \cite{Wyrowski1988} is expanded to all three spatial dimensions (3D-IFTA) and yields the phase offsets $\left\{\phi_j\right\}$ until a desired uniformity is reached. The deduced set of $\left\{\phi_j\right\}$, finally, completely determines the total phase-only transmission $T^{\text{tot}}$ for each spot placed in the working volume by $\left\{\mathbf{K}_j, f_j\right\}$. Examples for 3D-focus distributions (intensity cross sections) and corresponding phase modulations $\Phi^{\text{tot}}\left(\mathbf{r}\right)$ are depicted in Fig.\,\ref{fig:3dfoc}.
\par
\begin{figure}
    \centering
    \includegraphics[width=0.5\textwidth]{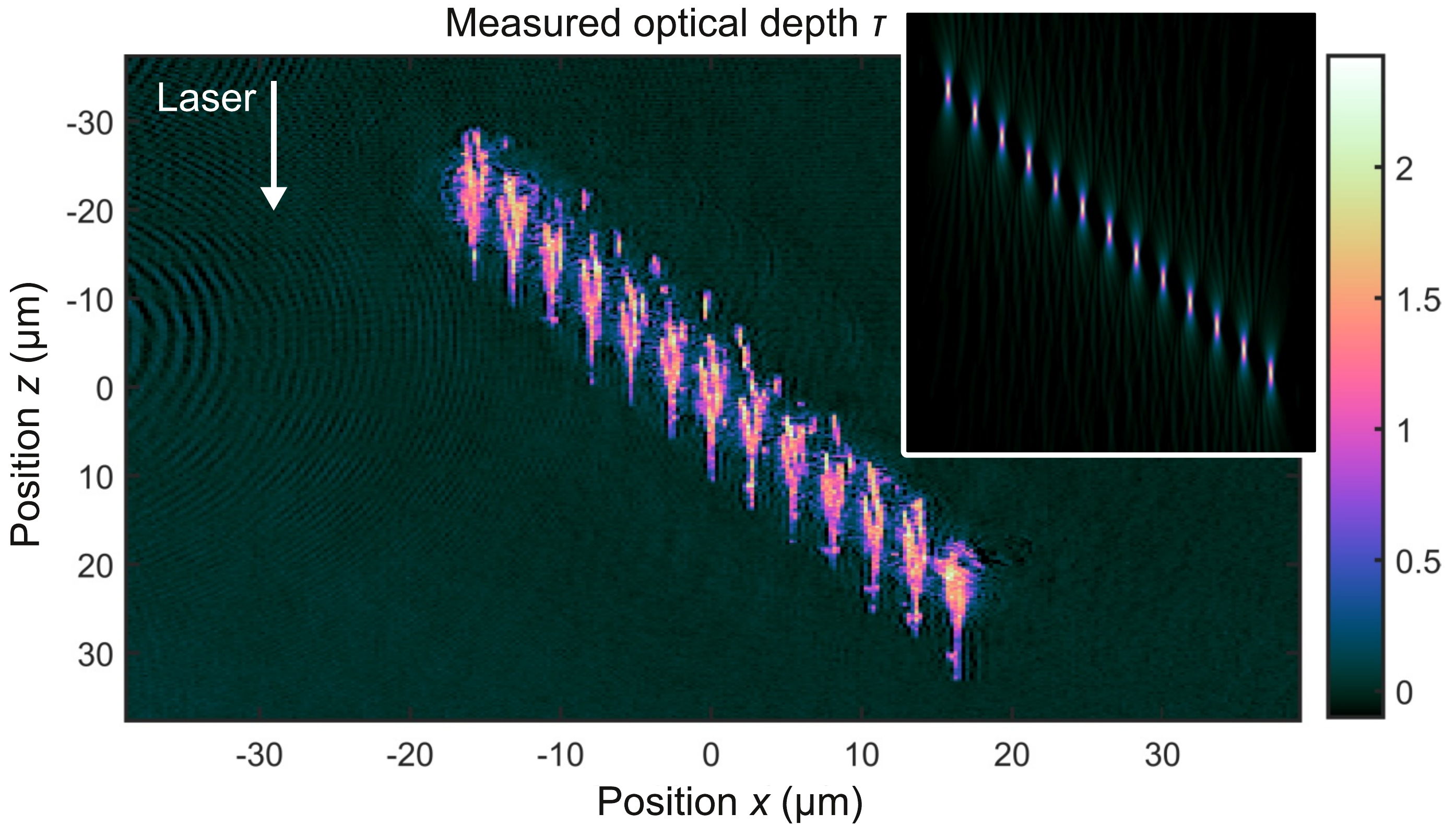}
    \caption{Measured optical depth $\tau\left(x, z\right)$ for a multi-spot focus distribution, see inset, inside the volume of borosilicate glass using transverse pump-probe microscopy \cite{Bergner2018, Jenne2018b}. The beam propagates from negative to positive $z$-direction. The absorption projection is composed of $13$ single zones and clearly indicates that energy is deposited spatially controlled by the beam splitting concept.\cite{Jenne2018b, Flamm2019}}
    \label{fig:scheiss2}
\end{figure}

 
We designed a diffractive optical element capable to split the illuminating fundamental Gaussian beam into $13$ copies, see Fig.\,\ref{fig:scheiss2}\,(a)\cite{Flamm2019}. Referring to the schematic shown in Fig.\,\ref{sschool}, the collimated raw beam illuminates the beam shaping DOE [Fig.\,\ref{sschool}\,(d)] which forms a $2f$-arrangement with the focusing unit [Fig.\,\ref{sschool}\,(e)]. Each focus exhibits a well defined longitudinal and transverse distance with respect to the respective neighbour, see simulated intensities in Fig.\,\ref{fig:scheiss2}\,(b) ($z$-axis corresponds to propagation direction). The successful energy deposition into the volume of a transparent workpiece is proven by transverse pump-probe microscopy, for details see Refs.\,\citenum{Bergner2018, Bergner2018a, Grossmann2016}. The measured optical depth $\tau\left(x, z\right)$ in non-strengthened Corning Gorilla$^\text{\textregistered}$ glass from a single laser burst with $4$ pulses (pulse energy of $\unit[30]{\upmu J}$) emitted from a TruMicro 2000 with a pulse duration of $\unit[5]{ps}$ is provided in Fig.\,\ref{fig:scheiss2}\,(c).
Here, $\tau\left(x, z\right)$ at a probe delay of $\unit[5]{ns}$ indicates the optical losses due to, e.g., absorption or scattering on a transient temporal scale. For each of the 13 foci a single absorption zone is observed starting at the geometrical focus and expanding in direction of the incoming laser pulse. The behavior well known from focusing single Gaussian beams into transparent materials\cite{Grossmann2016, Bergner2018} is multiplied here due to the beam splitting concept. Intriguingly, the spatial distribution of induced modification corresponds to the simulated focus positions and even at this small lateral and longitudinal spot separation of $\unit[3]{\upmu m}$ no shielding or inhomogeneities in between the individual spots is obtained. This denotes a precondition for advanced material processing in particular for scaling throughput in the field of welding, material functionalization or cutting.

\section{Laser surface structuring and percussion drilling}\label{sec:drill}
Drilling processes by ultrashort laser pulses meet the demand for high-end applications in the display and electronics industry. Especially the manufacturing of microstructures requires highest accuracy and minimal damage of the workpiece. Although the materials to be processed are very diverse---from highly absorbent to transparent\cite{chichkov1996femtosecond, furusawa1999ablation, shah2001femtosecond, Kumkar2016}---the process strategy is usually based on a Gaussian focus (single spots or multiple copies\cite{Kumkar2017, Flamm2019}) being directed over the workpiece using scanner optics or axis systems. Although, as mentioned, high quality machining results are usually obtained, industry requirements regarding efficiency and process speed often cannot always be met.  Additionally, a variety of applications, like the production of blind holes in multi-layer stacks or through holes in metal foils demand specific processing constraints. For example, the application of fine metal masks (FMMs) for the fabrication of OLED displays requires exact rectangular hole shape as well as tailored taper angles and minimized residual particle contamination. Decisive for a realization with ultrashort pulsed laser platforms, finally, is the achievable throughput.\cite{Kumkar2017, Flamm2019, kudryashov2019high, pavlov201910}.
\par
\begin{figure}
    \centering
    \includegraphics[width=0.5\textwidth]{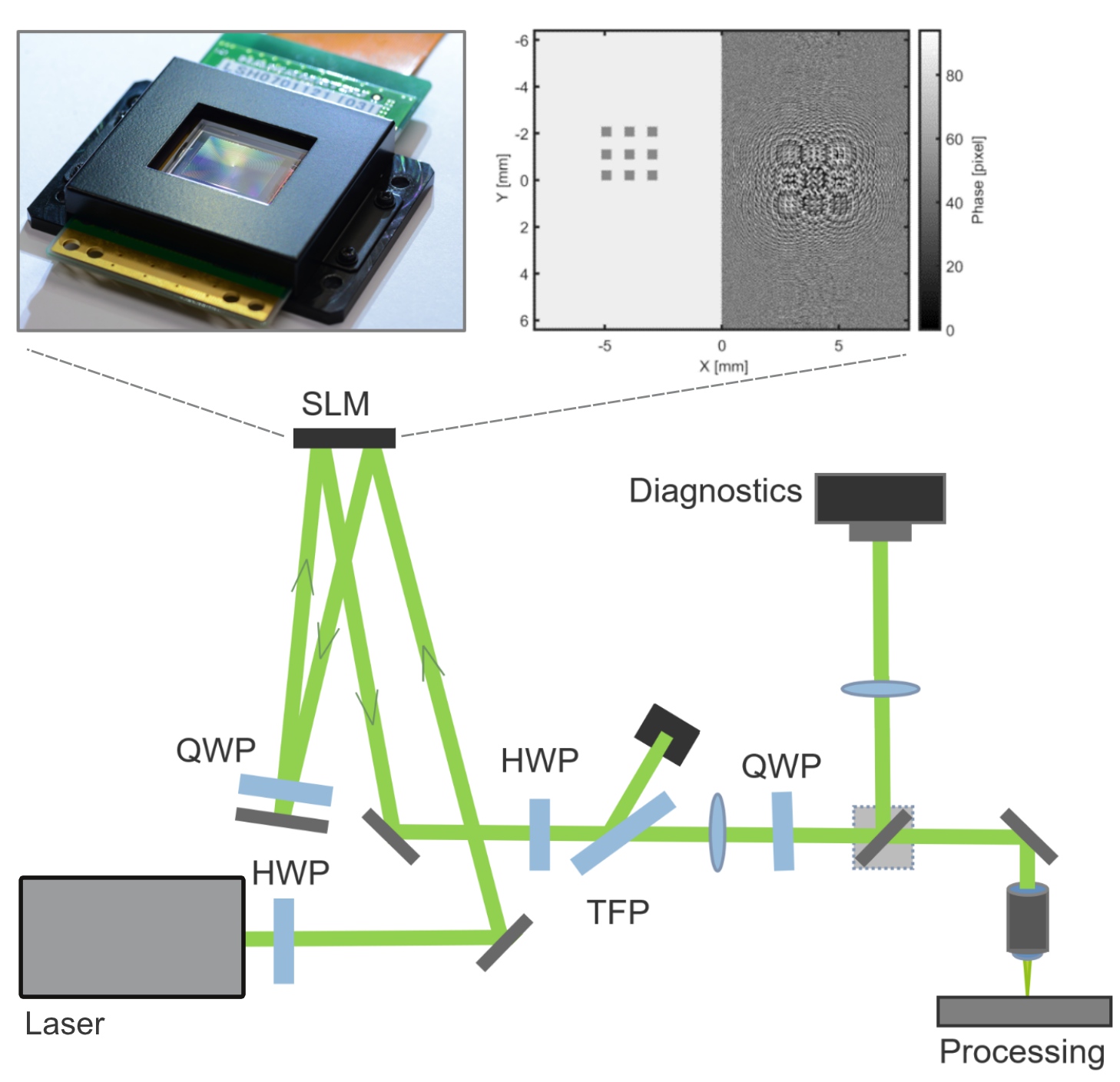}
    \caption{Optical setup for shaping multiple beam properties. Ultrashort laser pulses emerging from a TruMicro 2000 and externally frequency converted to a wavelength of $\unit[515]{nm}$ are illuminating the SLM in a double-pass configuration; half-wave plate (HWP), quarter-wave plate (QWP), thin-film polarizer (TFP). In this configuration, the SLM's left-hand side in combination with QWP and TFP acts as amplitude modulator and imprints and amplitude mask to the illuminating field \cite{Flamm2019}.}
    \label{fig:guenn1}
\end{figure}
\begin{figure}
    \centering
    \includegraphics[width=0.35\textwidth]{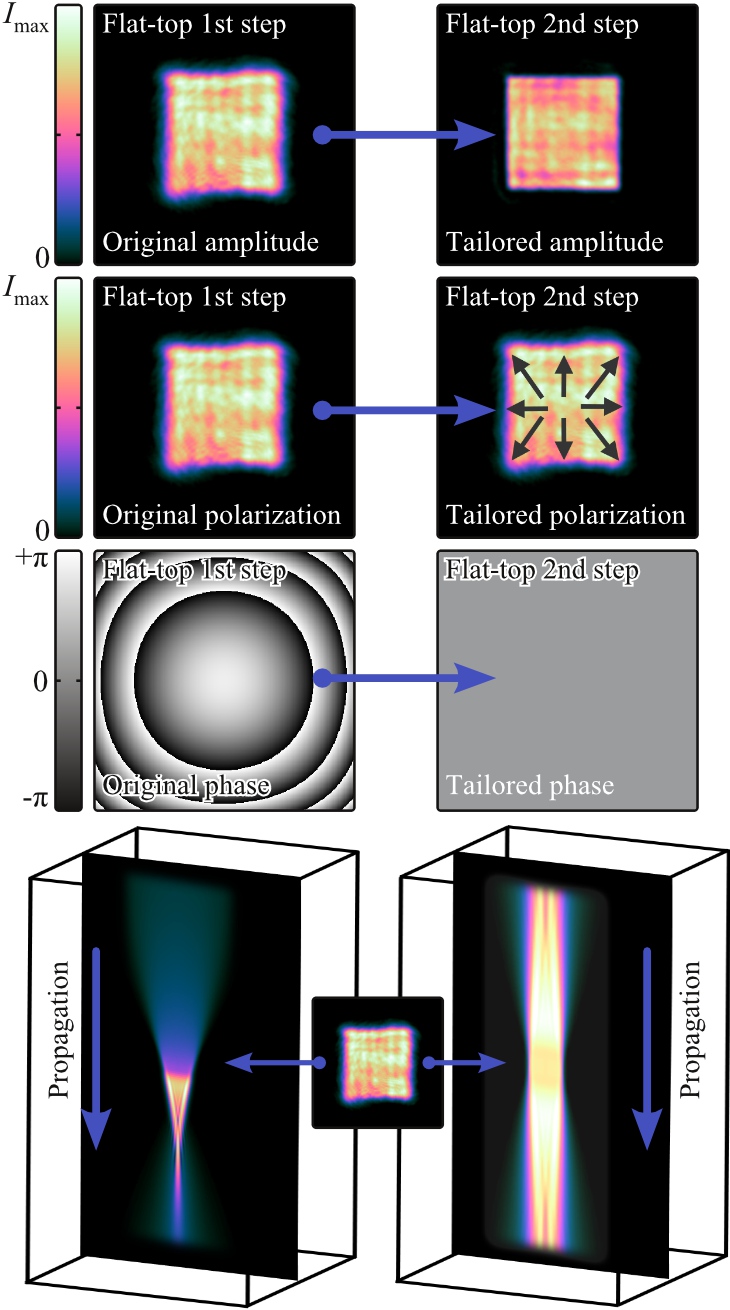}
    \caption{Two-step beam shaping of multiple beam properties. First step: phase-only shaping of flat-top beams.\cite{Dickey2000}. Potential second step: Tailored amplitude distribution using an amplitude mask or tailored polarization distribution using local quarter-wave or half-wave plates or tailored propagation properties based on phase manipulation. One possible setup to realize this concept is depicted in Fig.\,\ref{fig:guenn1} and employs a SLM as central beam shaping element.\cite{grossmann2020scaling}}
    \label{fig:guenn2}
\end{figure}

The optical concept described in the following, is realized with a liquid-crystal-on-silicon-based spatial light modulator (SLM) as central beam shaping element in a multi-illumination configuration, see Fig.\,\ref{fig:guenn1}.\footnote{For reasons of flexibility, this setup comprises an SLM. In consequence, the power performance is limited. A final processing optics may also consist of stationary beam-shaping components such as DOEs, ROEs or elements for geometric phase manipulation\cite{sakakura2020ultralow} mounted or fabricated on a single or multiple substrates and illuminated single or multiple times, cf.\,Fig.\,\ref{sschool}. Equally conceivable would be the use of unimorph deformable mirrors\cite{verpoort2020fast}. For the experiment conducted with the ultrashort pulsed laser operating at $\unit[257]{nm}$ wavelength, cf. Fig.\,\ref{fig:guenn8}, no SLM was used. Here, a setup with stationary beam shaping elements was developed.}
Starting from a conventional field-mapping concept for shaping a rectangular flattop distribution \cite{Dickey2000} (for alternatives, see Refs. \citenum{banas2014gpc, laskin2012variable, jiang2018liquid, kuang2015ultrafast, liu2018dynamic}), the SLM displays the required phase modulation at a well-defined subsection---in the present case the SLM's right hand side, see Fig.\,\ref{fig:guenn1}. The optical concept is designed in such a way that the flattop is generated at a second subsection of the SLM. This second beam shaping step enables to manipulate a further field property of the flattop, thus, in general, a well-defined spatial amplitude-, phase- or polarization distribution, see Fig.\,\ref{fig:guenn2}. By creating a plane phase distribution, for example, the propagation behavior can be manipulated to achieve maximum depth of field\cite{Pal2018, Scholes2020}. For the applications discussed in the following, such as laser drilling of thin foils, however, the amplitude distribution, in particular with regard to contour accuracy and edge steepness, will be decisive. Therefore, in the setup depicted in Fig.\,\ref{fig:guenn1}, the SLM's second illumination area in combination with quarter-wave plate (QWP) and thin-film polarizer (TFP) acts as amplitude modulator. In other words, the SLM imprints an amplitude mask to the illuminating optical field.
Without using a complex Fourier coding\footnote{The advantages in comparison to a classic phase-only IFTA\cite{Wyrowski1988} are in particular that no favorable initial transmission has to be found (an IFTA is only as good as its starting transmission) and that undesired diffraction orders and vortices are completely suppressed. We also see advantages in terms of homogeneity and edge steepness of the shaped profiles. The efficiency situation must be assessed on a case-by-case basis. Please note that for the present amplitude-mask concept local amplitude damping always comes at the expense of a direct reduction of available optical power at the workpiece.} such as iterative Fourier transform algorithms\cite{Wyrowski1988} almost any intensity profile can be generated. The required micrometer-scaled dimensions at the workpiece are finally achieved using the beam delivery system depicted in Fig.\,\ref{fig:guenn1} with adapted telescopic setups and middle- to high-NA focusing objectives.\cite{Flamm2019}.
\par
\begin{figure}
    \centering
    \includegraphics[width=0.30\textwidth]{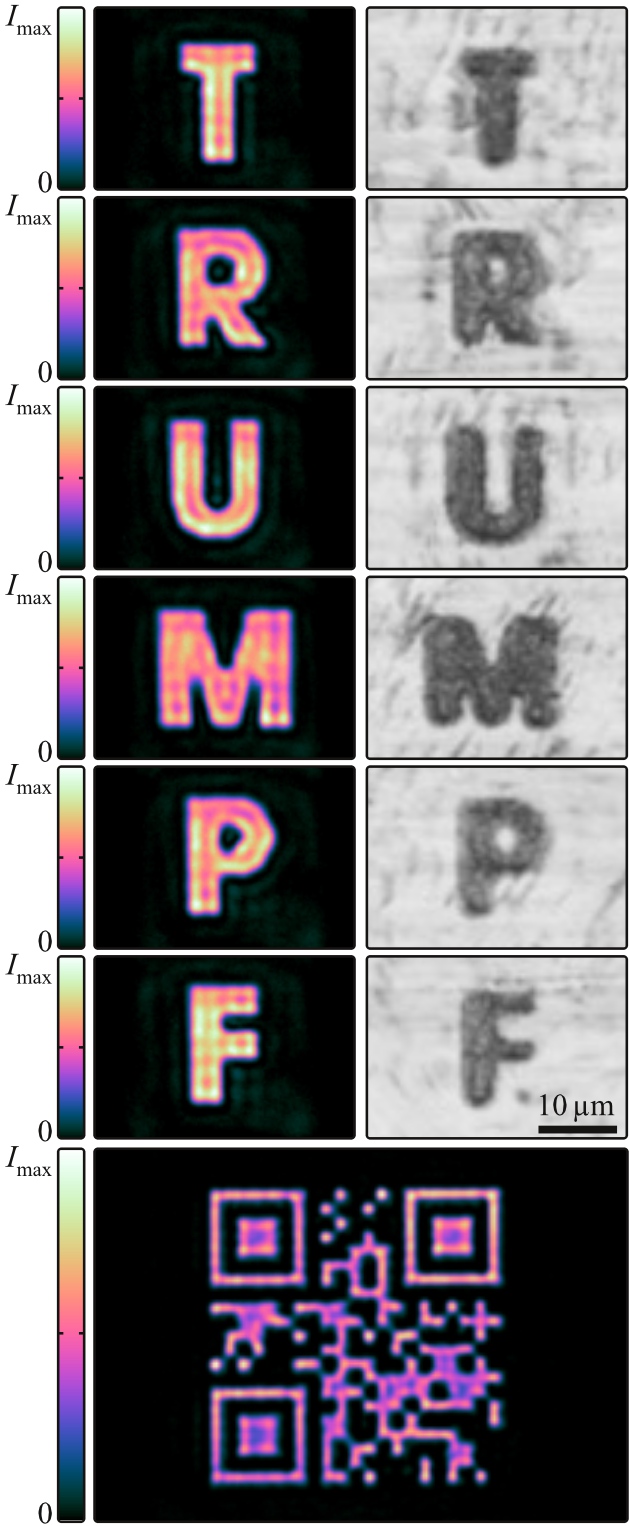}
    \caption{Direct marking of micrometer-scaled letters (microscope images, right column) from tailored intensity profiles (camera measurement, left column). The two-step beam shaping approach, cf. Fig.\,\ref{fig:guenn1}, enables simple realization of arbitrary amplitude distributions without applying Fourier coding techniques, see (and please scan) measured QR-code at the bottom.\cite{grossmann2020scaling}}
    \label{fig:guenn3b}
\end{figure}

\begin{figure*}
    \centering
    \includegraphics[width=0.8\textwidth]{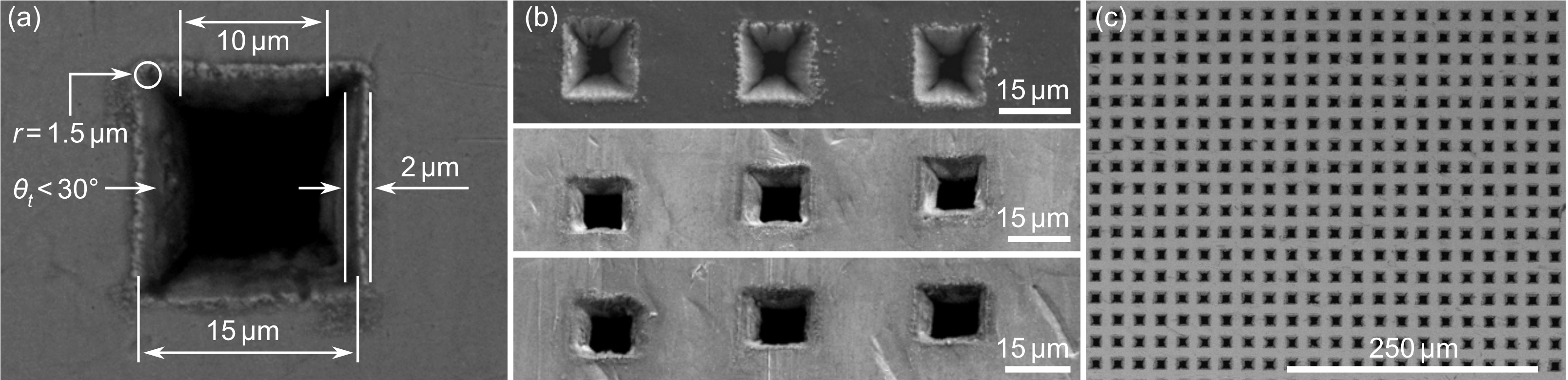}
    \caption{Drilling result using the flat-top beam profile depicted in Fig.\,\ref{fig:guenn6}. Direct drilling of quadratic-shaped holes in thin metal sheets with dimensions of $\sim\unit[10]{\upmu m}$ (a), different taper geometries and achieved hole density of $>\unit[1000]{ppi}$ (b), (c).\cite{grossmann2020scaling}}
    \label{fig:guenn6b}
\end{figure*}

 
Exemplary processing results of a micrometer-scale marking process and corresponding measurements of the intensity profiles are depicted in Fig.\,\ref{fig:guenn3b}. Here, letters of $\unit[10]{\upmu m}$-dimensions were directly marked onto the substrate's surface---no scanning system was required.
An other challenging application which can benefit from appropriate beam shaping flexibility is the percussion drilling of rectangular holes in thin metal foils (FMMs).  
Exemplary machining results with lateral dimensions below $\unit[20]{\upmu m}$ on an Invar foil with a thickness of $\unit[10]{\upmu m}$ are demonstrated in Fig.\,\ref{fig:guenn6b}. The accuracy regarding contour, edges and taper is a decisive factor of this process (a). Depending on the applied laser pulse parameter the exact geometry can be adjusted (b). With additional scaling by means of beam splitting, large scale samples with highly reproducible  holes can be achieved (c). Accordingly, the resulting pixel density or resolution, respectively can be $>\unit[1000]{ppi}$.
To achieve even further degrees of freedom, regarding the hole geometry, advanced beam shaping methods can be applied. In this context, the efficacy of the double stage shaping concept is demonstrated by means of controlled taper-angles for the drilled holes.
\begin{figure*}
    \centering
    \includegraphics[width=0.9\textwidth]{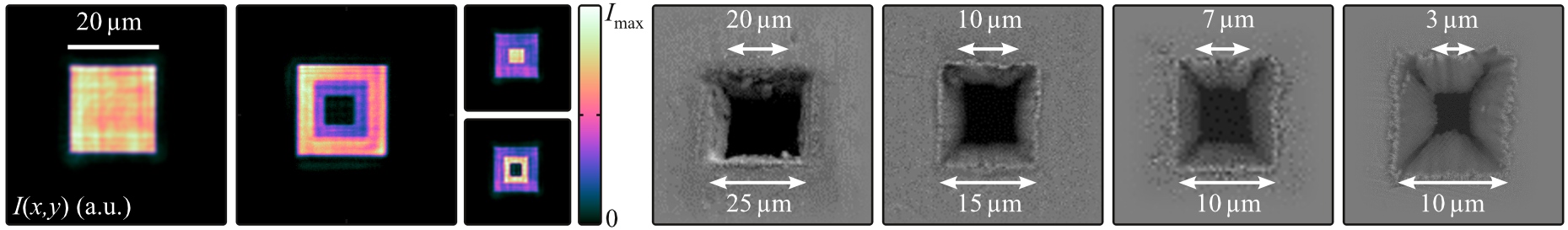}
    \caption{Controllable taper angles of drilled rectangular holes in invar foil with a thickness of $\unit[10]{\upmu m}$ for FMM applications. The additional adjustability is achieved using intensity graded flat-top-like beams on a $\upmu$m-scale at a wavelength of $\unit[515]{nm}$ (left-hand side).}
    \label{fig:guenn6}
\end{figure*}
Local intensity gradients of circularly-polarized flat-top-like beams enable to control the 3D hole geometry.  Figure\,\ref{fig:guenn6} shows SEM-images of details of laser-drilled holes in metal foils of $\sim\unit[20]{\upmu m}$-thickness with different entrance and exit hole dimensions. Please note, that the propagation behavior of the shown flat-top examples is similar due to approximately similar phase distributions, cf.\,Fig.\,\ref{fig:guenn2}. Taper angles are mainly controlled by the structured amplitude and resulting local intensity gradients.
\par 
The application of liquid-crystal-on-silicon-based SLMs shows several limitations regarding beam shaping performance and applicable laser parameters. Extensive diagnostic systems are provided to monitor the behavior of the beam shaping device during the application of high average powers. Using the telescopic setup, the SLM's second illumination step is imaged onto sensors for laser beam characterization, cf.~Fig.\,\ref{fig:guenn1}. Beam diagnostics may base on field reconstructions from caustic- \cite{merx2020beam, pang2020focal} or wavefront measurements\cite{sheldakova2007beam, Schulze2013} as well as from modal analysis\cite{flamm2013, pinnell2020modal}. In the simplest case a camera monitors the actual focus quality at a fixed $z$-position. In general, the flat-top's uniformity and contour accuracy will deviate from the ideal situation, cf. Fig.\,\ref{fig:guenn2}, for example, if a large number of mirrors is used for free-space beam delivery and wavefront aberrations exceed $\sim \lambda/10$ (p-v-error). Using the flexible SLM, the displayed phase and/or amplitude transmission can be adapted to the situation of the illuminating optical field. Simple control loops may compensate for potential wavefront aberrations using sets of Zernike modes\cite{Noll1976} or radial basis functions\cite{merx2020beam} until certain focus specifications, such as, e.g. homogeneity or edge steepness, are achieved. 
\par 
The phase-shifting performance of the SLM while exposed to high intensities or high average powers is usually only roughly known. To characterize the display spatially resolved interferometric techniques\cite{beck2010application} are used as well as global measurements of the diffraction efficiency. Possible applied laser parameters (wavelength, pulse durations, repetition rate, burst modes, etc.), however, are  manifold requiring a spatially resolved \textit{in situ} characterization technique. To achieve best beam shaping quality we calibrate the complete phase shifting performance of each pixel by spatially resolved polarization measurements.
\par
\begin{figure}
    \centering
    \includegraphics[width=0.48\textwidth]{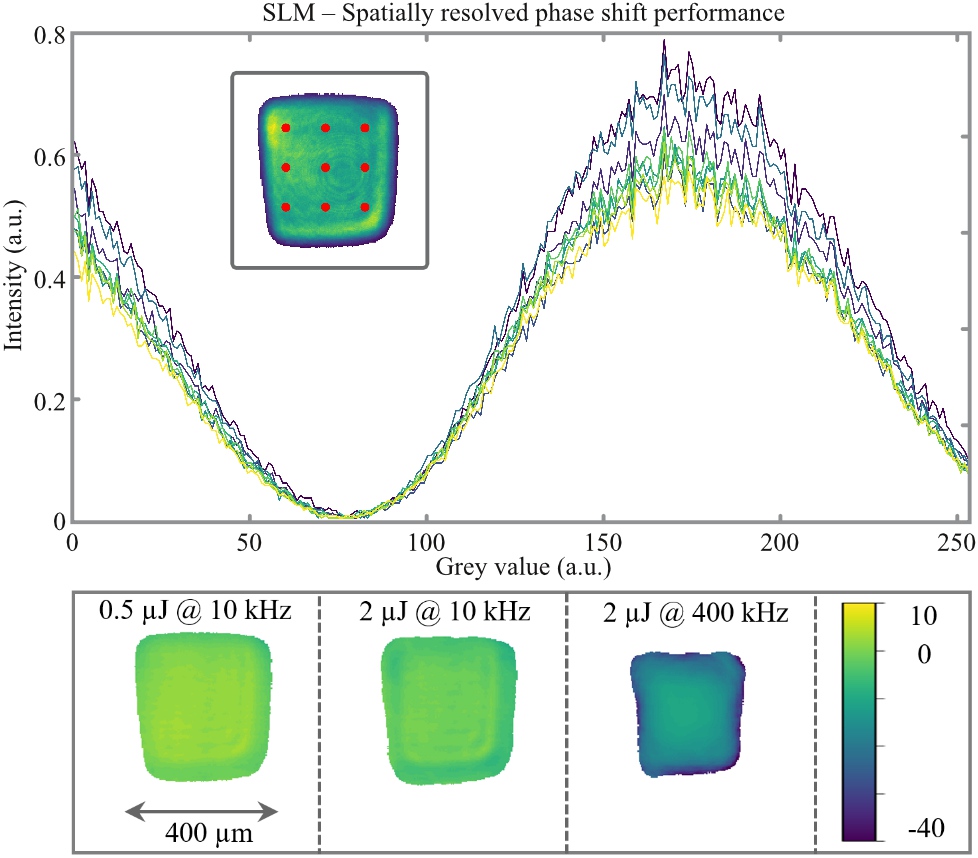}
    \caption{\textcolor{black}{Novel SLM characterization technique for \textit{in situ} measurement of the phase-shifting performance during display illumination with high optical powers or energies. Based on the setup depicted in Fig.\,\ref{fig:guenn1}, the beam shaping performance with pixel-resolved phase-shift-performance measurement at different pulse parameters is optimized. This method enables a direct measurement of necessary laser parameter-depended phase-correction masks for high quality beam shaping applications. The bottom row depicts exemplary phase correction maps (in arbitrary grey value units) for varying laser parameters.}}
    \label{fig:guenn7}
\end{figure}
\begin{figure}
    \centering
    \includegraphics[width=0.5\textwidth]{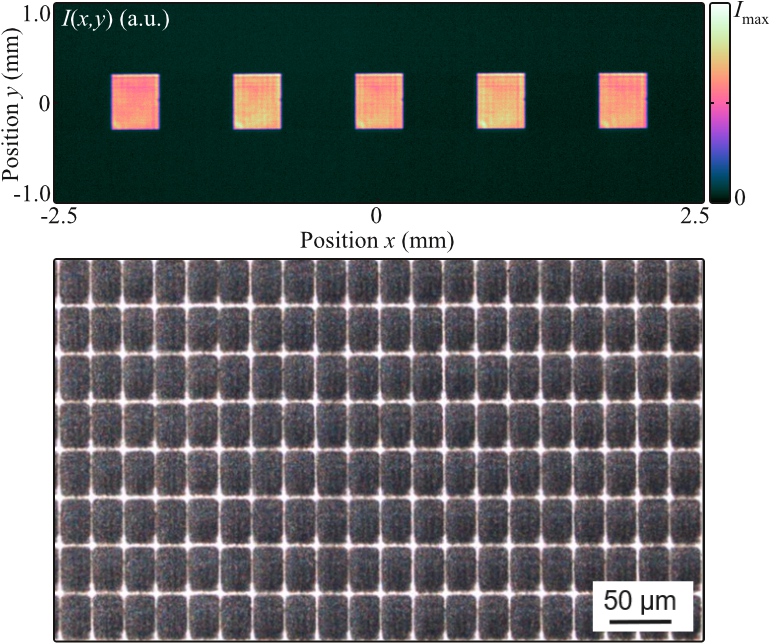}
    \caption{Splitting a flat-top focus distribution into multiple copies for parallel processing (top). The employed laser providing ps-pulses and operating at $\unit[257]{nm}$ wavelength is realized by an externally frequency converted TruMicro 5270. The processed silicon wafer exhibits rectangularly shaped patterns of $\unit[20]{\upmu m}$ dimensions (bottom) fabricated with fluences of $\sim\unit[1]{J/cm^2}$ showing the potential for drilling of masks and single-shot $\upmu$LED laser lift off, respectively.}
    \label{fig:guenn8}
\end{figure}

By irradiating the SLM with $45^\circ$ linear polarization and neglecting constant phase shifts the polarisation of the resulting electrical field $\mathbf{E}$ can be written as,
\begin{equation}\label{eq:pol1}
  \mathbf{E}(x,y) = \frac{1}{2} \begin{pmatrix} \sin \left(k d \left(n_{\text{e}}\left(x,y,U\right)-n_{\text{o}}\right)\right) \\ \imath \cos\left(k d\left(n_{\text{e}}\left(x,y,U\right)-n_{\text{o}}\right)\right) \end{pmatrix},
\end{equation}
with $k=2 \uppi/\lambda$ and the thickness of the liquid crystal layer $d$. The local polarization depends on the liquid crystal refractive index on both axis $n_{\text{e}}$ and $n_{\text{o}}$, respectively. In consequence, an applied $8$-bit pixel value corresponds to a specific voltage $U$ which changes the local phase retardation. 
By evaluating the intensity in one polarization state for each pixel, the exact phase modulation for each applied voltage can be calibrated.
With these measurements we can assess the necessary phase correction for each pixel at different laser parameters. In Fig.\,\ref{fig:guenn7}, the measured phase performance for selected SLM pixels is shown. With moderate laser parameters different pixels primarily exhibit a relative phase shift to each other which can be corrected directly by the spatially resolved measurement. These correction masks are shown in the corresponding graphic for different laser parameters, see bottom of Fig.\,\ref{fig:guenn7}. Thereby a substantial increase in phase shifting error for higher average laser powers can be observed. Especially with an increased thermal load the local additional phase shift becomes prominent. However, as long as the measured intensity profile over all pixel values is just shifted, a direct correction for beam shaping applications is possible. With a further increased laser power, additional errors start to occur, which can no longer be compensated by a simple pixel shift correction and indicate an approach to the damage threshold of the device.
\par 
In specific applications, which require highest power or short wavelength, beam shaping has to be realized with more suitable components. With an appropriate optical setup this can also be realized with multiple optical components. In Fig.\,\ref{fig:guenn8}, a measurement of an exemplary optimized beam shape is shown generated at a wavelength of $\lambda = \unit[257]{nm}$. The designed setup similar to the double illumination concept of Fig.\ref{fig:guenn1} was realized solely with static and wavelength-suitable components. Here, no SLM was used, since conventional liquid crystal displays show significant absorption below $\unit[400]{nm}$ wavelength. To demonstrate the capabilities of beam shaping this ``exotic'' laser parameter an additional microscope image of a processed silicon wafer is shown. This workpiece was processed with several beam split flat-top focus distributions of $\unit[20]{\upmu m}$ dimensions and $\sim\unit[2]{ps}$ ultra short laser pulses at $\lambda = \unit[257]{nm}$. We employed an externally frequency-converted TruMicro 5270 laser based on the established industrial ultrafast laser platform TruMicro Series 5000\cite{Quentin2020, Haefner2021}.\par
By using a rectangular-confined spatial energy distribution instead of a round-shaped Gaussian focus with a slowly decaying amplitude, rectangular patterns can be realized that are used in a variety of applications like drilling of masks or for example laser lift off. Here, the combination of energy absorption in the DUV regime paired with ultra-short temporal energy deposition, enables a minimum thermal damage to functional layers made of, e.g., isolators or semiconductors. Adapting the flat-top dimensions to the feature size of the semiconductor and applying beam splitting, cf.~Fig.\,\ref{fig:guenn8}, parallel processing, single-shot lift-off strategies can be realized that meet the throughput demands of industrial production lines. \par 
\textcolor{black}{The broad accessibility of liquid-crystal displays makes their use in industrial-grade processing optics more and more conceivable. The steadily increasing refresh rates up to several hundred Hertz may even allow the display's use as adaptive components during the actual machining. For example, too much energy is usually deposited at the reversal points of a scanning process for ablation or marking applications. One could use the SLM to switch the beam shape during scanning to adapt intensities and to achieve particularly homogeneous processing results. Especially here, \textit{in situ} diagnostics of the phase-shifting performance is of importance, as demonstrated by the direct measurement of pulse-energy dependent phase correction masks, cf.~Fig.\,\ref{fig:guenn7}. For the further industrial dissemination of adaptive beam shaping concepts based on liquid-crystal displays---not only with regard to laser materials processing---, the effort in calculating the required phase masks needs to be reduced. Usually, complex Fourier algorithms are used for this purpose which require advanced knowledge in the field of wave optics. We would like to emphasize once again, that using the presented double-illumination concept, cf.~Fig.\,\ref{fig:guenn1}, where the SLM acts as both phase and amplitude modulator, no sophisticated phase-coding algorithms are necessary.}

\section{Conclusion}
Structured light concepts as enabler for advanced ultrafast laser materials processing have been demonstrated. Based on selected challenging application examples, such as cutting of glass tubes or drilling of fine metal masks, relevant for medical application and consumer electronics, respectively, we reviewed the generation and beneficial use of customized focus distributions. Several optical concepts were introduced for shaping a variety of beam classes ranging from tailored non-diffracting beams to adapted 3D-beam splitters operating in the near infrared down to the deep ultraviolet. As a matter of principle, concepts were examined that can withstand high average and peak powers showing potential to be central parts of future processing optics forming ultrafast focus distributions with tens-of-millijoules pulse energies and kilowatt-class average powers. \textcolor{black}{Finally, conceptual solutions were discussed in which adaptive optics based on liquid-crystal displays can act as central beam shaping elements in future industrial-grade processing optics.}

\subsection*{Disclosures}
The authors declare no conflicts of interest.



\footnotesize
\bibliography{report}   
\bibliographystyle{spiejour}   




\end{spacing}
\end{document}